\documentclass[10pt,journal]{IEEEtran}
\usepackage{fontspec}
\usepackage{amsmath,amssymb}
\usepackage{unicode-math}
\usepackage{graphicx}
\usepackage{booktabs,tabularx,array}
\usepackage{microtype}
\usepackage[hidelinks,breaklinks=true]{hyperref}
\providecommand{\tightlist}{\setlength{\itemsep}{0pt}\setlength{\parskip}{0pt}}
\newcolumntype{Y}{>{\raggedright\arraybackslash}X}
\hypersetup{pdftitle={Agentic AI for Staged Three-Dimensional Finite-Element Tunnel Modelling},
  pdfauthor={Ochok Duangsano; Kuo-Chieh Chao; Sompote Youwai; Nattavich Sittiamornporn; Chana Phutthananon; Pornkasem Jongpradist}}
\begin{document}
\title{Agentic AI for Staged Three-Dimensional Finite-Element Tunnel Modelling}
\author{Ochok~Duangsano, Kuo-Chieh~Chao, Sompote~Youwai, Nattavich~Sittiamornporn, Chana~Phutthananon, and Pornkasem~Jongpradist%
\thanks{O. Duangsano is with the Asian Institute of Technology, Pathum Thani, Thailand, and with Geotechnical and Foundation Engineering Company Limited, Bangkok, Thailand.}%
\thanks{K.-C. Chao is with the Asian Institute of Technology, Pathum Thani, Thailand.}%
\thanks{S. Youwai, C. Phutthananon and P. Jongpradist are with the AI Research Group, Department of Civil Engineering, King Mongkut's University of Technology Thonburi (KMUTT), Bangkok, Thailand. Corresponding author: S. Youwai (e-mail: sompote.you@kmutt.ac.th).}%
\thanks{N. Sittiamornporn is with GOE Consultants Company Limited, Bangkok, Thailand.}}
\markboth{Duangsano \MakeLowercase{\textit{et al.}}: Agentic AI for Staged 3D Finite-Element Tunnel Modelling}{}
\maketitle
\begin{abstract}
Three-dimensional finite-element (FE) analysis gives the most complete mechanical picture of tunnelling-induced ground movement. It is also slow. An experienced engineer must drive the graphical interface of PLAXIS 3D through project set-up, stratigraphy, materials, structural elements, meshing and staged calculation. This paper presents a pipeline in which a large language model (LLM) drives PLAXIS 3D directly through the Model Context Protocol (MCP), with no human operating the interface. A purpose-built MCP server exposes the official remote-scripting interface as 19 typed tools. Four agents share the work: an orchestrator, a geometry agent, a calculation agent and a verification agent. They draw on 15 inspectable domain-knowledge skill modules and a provenance-tagged intake stage. The verification agent is information-barriered. It audits every model against the raw specification at a pre-mesh checkpoint and again as a gate on the calculation. The framework was evaluated on 12 single-tunnel problems in Bangkok subsoil across four capability tiers, using a 15-arm ablation matrix. Thirteen automated arms and a manual expert baseline were executed, with one run per cell. The full system built 12 of 12 models at a mean of 6.1 min with a clean audit on every one. The same models took 24.6 min by hand. Domain skills were decisive. With the complete architecture but all 15 modules withheld, no model was produced. Base-model capability acted as a threshold rather than a gradient. Fable 5, Opus 5 and Sonnet 5 each built 12 of 12, and Haiku 4.5 halted on all 12. Among models that built, agreement was exact. 107 of 111 comparable cells reproduced the reference mesh to the element. 83 of 86 solved cells agreed with the hand-built expert model within 1.01 \% in maximum settlement and 1.21 \% in trough-width parameter. Every exception traced to a value misread at intake. The reliability problem in agentic FE automation therefore lies in specification reading, not in code generation.
\end{abstract}
\section{Introduction}\label{introduction}

Shield tunnelling beneath dense urban areas induces ground movement. Predicting the surface settlement trough is a central problem in urban geotechnical engineering. The problem is acute in cities founded on thick soft clay. Bangkok is the canonical case, with its expanding MRT network bored through the soft and stiff Bangkok clays. There, settlement prediction governs building-damage assessment, alignment selection and the specification of tunnel-boring-machine (TBM) operating pressures. Prediction methods range from empirical trough relations (Peck, 1969) and closed-form solutions (Loganathan and Poulos, 1998) to data-driven models (Suwansawat and Einstein, 2006). Three-dimensional FE analysis in a package such as PLAXIS 3D gives the most complete mechanical representation. It can model ring-by-ring TBM advance, face and grouting pressures, shield conicity, lining installation, the groundwater regime, and layered soil behaviour with constitutive models calibrated for Bangkok subsoil (Surarak et al., 2012).

That fidelity costs expert labour. A staged 3D model is assembled through a fixed sequence: project set-up, stratigraphy, material sets, structural elements, meshing, staged calculation and result extraction. The sequence is usually executed through the graphical interface over hours to days. The process is error-prone. Unit and sign conventions, reference elevations and phase dependencies are all easy to get wrong. The process is hard to reproduce across analysts. It also places a scarce expert on the critical path of every parametric study. The remote-scripting Python API moves this effort rather than removing it. The same expert must still write, debug and adapt the scripts for each problem.

Two developments make full automation plausible for the first time. LLMs have progressed from code assistants to agents that plan multi-step workflows and drive external tools. MCP standardises how such agents connect to external software (Anthropic, 2024). A simulation package can therefore be exposed to any MCP-capable platform through one server. Yet no published system autonomously constructs, verifies, solves and post-processes staged 3D shield-tunnelling models in a commercial geotechnical FE package. Nor have the causes of reliability in agentic simulation been isolated experimentally. The candidate causes are architecture, domain knowledge and base-model capability.

Figure 1 sets the two workflows side by side against the same engineering question. On the left, the engineer interprets the brief and drives all ten modelling stages by hand. There is no separation of roles and no independent check. On the right, the engineer states the problem once, confirms any value the intake stage flags as assumed, and returns only at sign-off. Between those points nobody operates the interface. An orchestrator sequences a modelling agent and a calculation agent. A verifier sits outside that group. It receives the specification directly and returns a verdict, and it does not share the builders' context. All four agents draw on the same skill modules. The only route to PLAXIS 3D is the MCP server. The claim is not that the human disappears. The claim is that the human leaves the modelling loop, and that what replaces them is inspectable.

\begin{figure*}[!t]
\centering
\includegraphics[width=\textwidth,height=0.48\textheight,keepaspectratio]{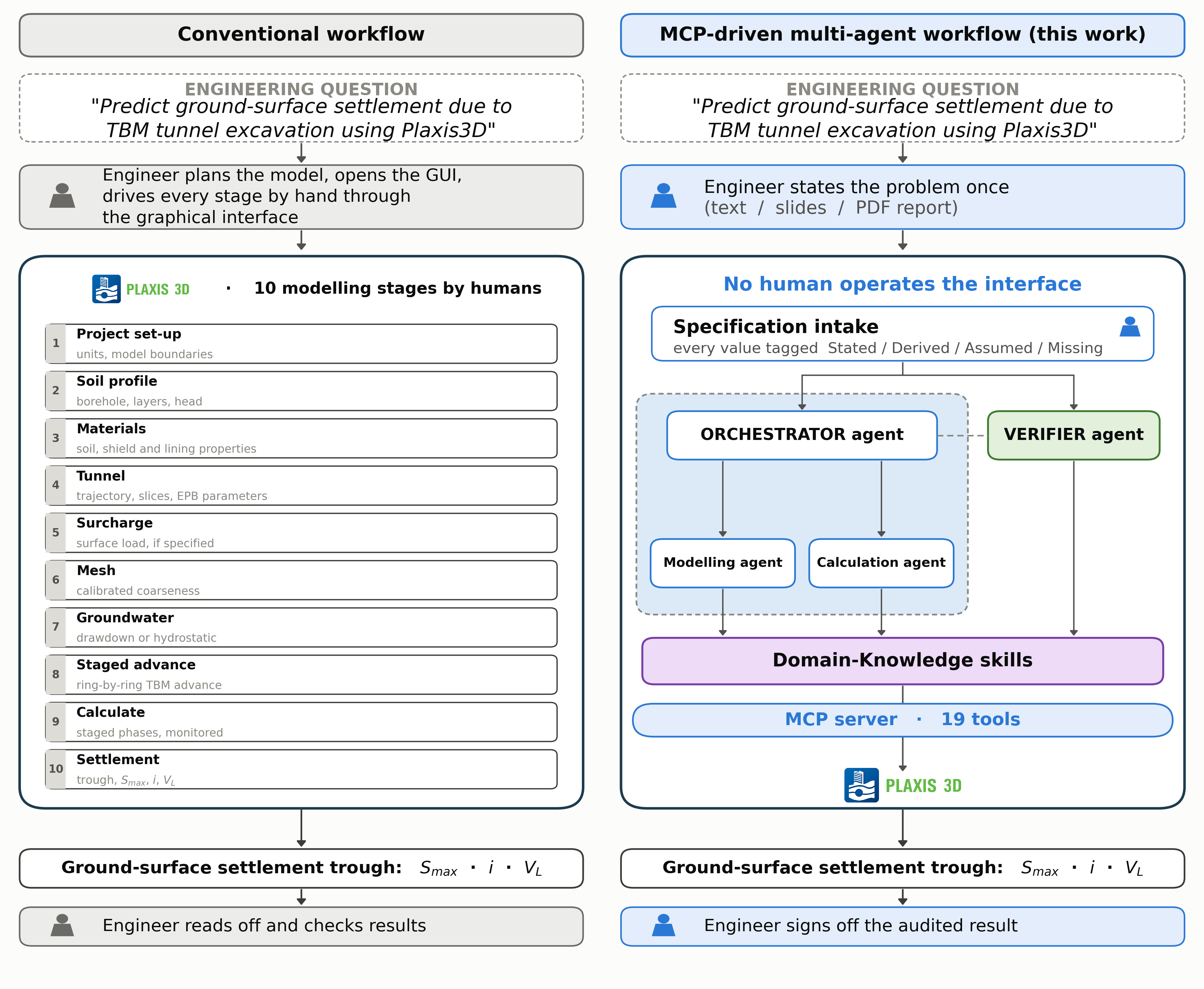}
\caption{Conventional expert-driven workflow (left) and the MCP-driven multi-agent workflow of this paper (right).}
\label{fig:1}
\end{figure*}

The contributions are as follows. (1) A PLAXIS 3D MCP server that exposes the official remote-scripting interface as 19 MCP tools. (2) An orchestrated four-agent architecture with an information-barriered, read-only verification agent. This agent re-derives the model from the raw specification, checkpoints the geometry before meshing, and gates every calculation. (3) Fifteen inspectable and ablatable domain-knowledge skill modules with a provenance-tagged intake stage. (4) A tiered single-tunnel benchmark in Bangkok subsoil and a 15-arm, 180-case ablation matrix. The matrix spans architecture variants, verification removal, skill removal, four LLM capability rungs and cross-vendor portability. (5) Empirical findings on what makes agentic FE automation reliable.

\section{Related Work and Research Gap}\label{related-work-and-research-gap}

\textbf{Prediction of tunnelling-induced ground movements.} The empirical tradition begins with Peck (1969). He observed that the transverse settlement trough is well approximated by an inverted Gaussian curve. The curve is characterised by the maximum settlement \(S_{max}\) and the trough-width parameter \(i\):

\[S_v(x) = S_{max} \exp\left(-\frac{x^2}{2i^2}\right) \tag{1}\]

The trough volume per metre of advance \(V_s\) and the ground-loss ratio \(V_L\) follow as

\[V_s = \sqrt{2\pi}\, i\, S_{max}, \qquad V_L = \frac{V_s}{\pi D_s^2/4} \tag{2}\]

where \(D_s\) is the excavated diameter. The trough width is estimated from the tunnel-axis depth \(Z\) as

\[i = kZ \tag{3}\]

O'Reilly and New (1982) take \(k \approx 0.5\) for clays and \(k \approx 0.25\) for granular soils. Mair and Taylor (1997) give \(k\) = 0.4 to 0.6 in clay and 0.25 to 0.45 in granular materials. Bangkok field data follow this pattern. Duangsano et al.~(2023) report \(k\) = 0.4 to 0.6 for clayey and mixed-face ground and 0.3 to 0.5 where the tunnel is driven entirely in sand. Figure 2 defines these quantities over a typical Bangkok profile. \(V_L\) is taken at the excavation boundary rather than at the lining.

\begin{figure*}[!t]
\centering
\includegraphics[width=\textwidth,height=0.48\textheight,keepaspectratio]{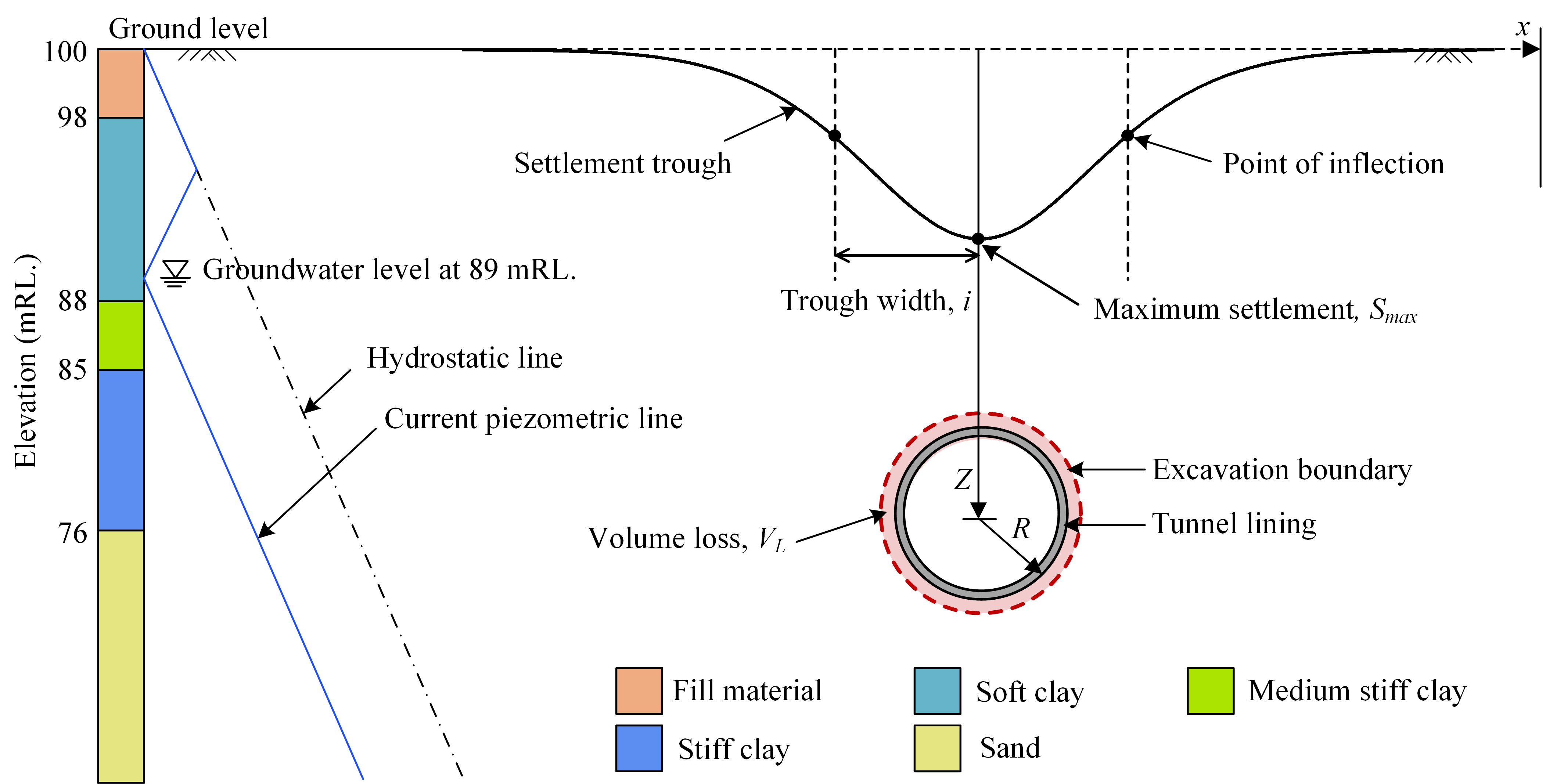}
\caption{Settlement trough (Peck, 1969) and a typical Bangkok soil profile.}
\label{fig:2}
\end{figure*}

Analytical solutions (Sagaseta, 1987; Verruijt and Booker, 1996; Loganathan and Poulos, 1998) are fast and transparent. They cannot represent layered stratigraphy, advanced constitutive behaviour, drawdown or staged EPB mechanics. Numerical simulation fills that gap. Surarak et al.~(2012) calibrated Hardening Soil parameters for Bangkok clays. Likitlersuang et al.~(2014) proposed simplified FE procedures benchmarked against Blue Line data. Lueprasert et al.~(2017) found that shell elements combined with a solid grouting layer best reproduced the measured trough. Phutthananon et al.~(2023) modelled the staged EPB advance to study pile response to twin tunnelling. Related Bangkok studies compare simulation models of differing complexity for tunnel-structure interaction (Heama et al., 2021; Lueprasert et al., 2023; Ekaksorn et al., 2024; Sittiamornporn et al., 2024). Their common limitation is not mechanical fidelity but labour. Every model is hand-built by an expert. Machine-learning methods (Suwansawat and Einstein, 2006; Xu et al., 2024) predict settlement but do not produce the underlying mechanical model. None of this literature treats construction of the model itself as the object of automation.

\textbf{LLM agents for computational engineering.} MCP standardises agent-to-tool connectivity through host, client and server roles with machine-readable schemas. It has been applied to scientific computing (Pan et al., 2025), traffic simulation (Ye et al., 2025) and BIM interaction (Heimig-Elschner et al., 2026). On this substrate, FeaGPT (Qi et al., 2025) implements a geometry, mesh and simulation pipeline over CalculiX. VFEAgent (Zhang et al., 2026) adds multimodal intake with integrated verification. Geng and co-workers generate SAP2000, ETABS and OpenSees analyses from natural language (Geng et al., 2026a, 2026b, 2026c). Foam-Agent 2.0 (Yue et al., 2025) and ChatCFD (E. Fan et al., 2025) automate OpenFOAM with 88.2 \% and 82.1 \% success. Closest to this study, Elbasani et al.~(2026) drive AutoCAD 2024 and ETABS v21 through two MCP servers exposing 24 and 18 tools. They show that protocol overhead is negligible, at 45 to 120 ms per tool call. Two differences matter. Their pipeline has no staged construction: no soil continuum, no constitutive-model selection, and no carry-forward of stress state between phases. Their validation metric is the geometric fidelity of node coordinates plus a code-check pass or fail. It is not the convergence and physical plausibility of a nonlinear solution. Two lessons recur in this literature. First, decomposition with explicit verification is generally reported to improve reliability. Hallucinated parameters and silently wrong models are the dominant failure mode (Geng et al., 2026c; Zhang et al., 2026). Liang et al.~(2025) report a single-agent system that failed all ten trials on a task their two-agent system completed reliably. Section VI-B reports a case where that lesson does not transfer. Second, evaluation is usually restricted to end-to-end success rates, so the causes of reliability are rarely isolated.

\textbf{LLMs in geotechnical engineering.} Three 2026 reviews frame the field. Sheil et al.~(2026) expect AI to remain a tool with a human in the loop. They note that AI models ``may lack a similar mechanism for validation'' to the cross-checking engineers perform routinely. Fan et al.~(2026) argue that domain adaptation is not optional and that reproducibility is the binding constraint. Chatterjee et al.~(2026) position LLMs as ``intelligent co-pilots rather than autonomous designers''. Among systems, GeoSim.AI (Bekele, 2025) assists a human scripting session. A generative-AI and MATLAB pipeline automates numerical analysis and synthetic data generation (Parsa-Pajouh, 2025). GeoLLM (Xu et al., 2025) targets design-code calculation. GeoMCP (Bekele, 2026) combines MCP with method cards and a symbolic engine but restricts itself to analytical methods. Youwai et al.~(2026) compare router-based architectures for foundation design. Hu et al.~(2025) deploy a six-agent assistant on a TBM control system that advises rather than models. Every geotechnical system sits at the assistant or orchestrator level. None provides the independent check the reviews identify as missing.

\textbf{Research gap.} Five gaps follow. \textbf{G1:} there is no end-to-end agentic automation of 3D geotechnical FE modelling. Existing agents target single-physics problems far lighter than staged shield tunnelling. \textbf{G2:} there is no independent, information-barriered verification. Where verification exists, it is a self-check run in the context that built the model, so a misread specification corrupts both. \textbf{G3:} there is no provenance-tagged specification intake, although specification misreading is a dominant failure mode. \textbf{G4:} no component-level ablation isolates orchestration, specialisation, verification, knowledge and base-model capability on one problem set. \textbf{G5:} domain knowledge is implicit. It is embedded in prompts or weights, where it cannot be inspected, versioned or removed.

Table I sets this work against the closest systems. Five differences follow. (i) The target problem class is staged 3D shield tunnelling in a commercial geotechnical package. It includes Bangkok Hardening-Soil stratigraphy, three groundwater regimes, and TBM pressures with their sign and reference conventions (G1). (ii) Verification is an information-barriered gate rather than a self-check. It is invoked twice and enforced at the artefact level, because the audit configuration is transcribed from the source documents rather than from the builder's configuration (G2). (iii) Intake is provenance-tagged (G3). (iv) The architecture is measured rather than asserted, through the 15-arm ablation (G4). (v) Domain knowledge is held as inspectable, ablatable files (G5).

\begin{table*}[!t]
\caption{Comparison with the closest related frameworks.}
\label{tab:1}
\centering\footnotesize
\setlength{\tabcolsep}{4pt}\renewcommand{\arraystretch}{1.15}
\begin{tabularx}{\textwidth}{YYYYY}
\toprule
Framework & Domain / software & Architecture & Verification & Ablation \\
\midrule
GeoMCP (2026) & Analytical geotechnics (EC7) & Orchestrator + symbolic engine & Dimensional checks & No \\
GeoSim.AI (2025) & Geomechanics simulators & Single assistant, human in loop & None (human) & No \\
FeaGPT (2025) & General FEA (CalculiX) & Agentic pipeline & Self-validation & No \\
VFEAgent (2026) & General FEA & Multi-agent & Same context & Partial \\
Geng et al. (2026a–c) & Structural frames & Multi-agent, two-stage & Self-correction & Partial \\
Foam-Agent 2.0 (2025) & CFD (OpenFOAM) & Composable multi-agent & Error-correction & No \\
Elbasani et al. (2026) & CAD + structures & Two MCP servers, one client & Code-compliance self-check & No \\
Hu et al. (2025) & TBM operations & 6 agents, no MCP & None (human) & Module-level \\
\textbf{This work} & Staged 3D shield tunnelling (PLAXIS 3D) & Orchestrated 4-agent + 15 skills over MCP & Independent, barriered audit: checkpoint + gate & 15 arms × 12 problems \\
\bottomrule
\end{tabularx}
\end{table*}

\section{The PLAXIS 3D MCP Server}\label{the-plaxis-3d-mcp-server}

MCP is an open standard for connecting AI applications to external systems. It uses a client-server architecture (Fig. 3). A \textbf{host} is where the user interacts with the AI application. It manages permissions and routes requests. A \textbf{client} inside the host maintains a one-to-one connection with one server. A \textbf{server} exposes tools, data or services over standardised interfaces.

\begin{figure*}[!t]
\centering
\includegraphics[width=\textwidth,height=0.48\textheight,keepaspectratio]{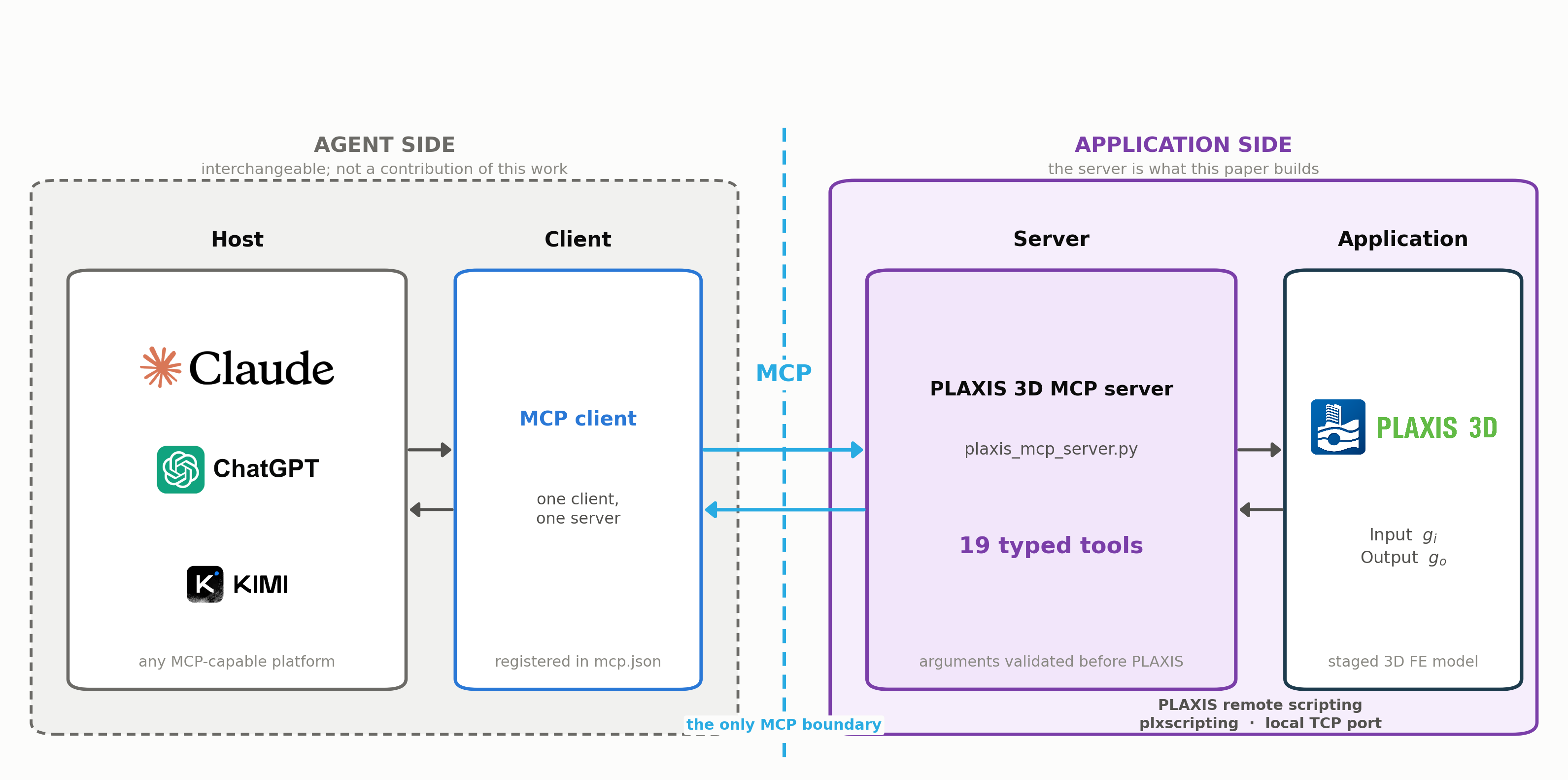}
\caption{MCP architecture.}
\label{fig:3}
\end{figure*}

The separation matters here. The protocol, not the platform, defines the contract, so one server serves every host that speaks it. The same \texttt{plaxis\_mcp\_server.py} is driven unchanged from Claude Code and from GPT-Codex. That is what makes the portability arms a like-for-like comparison rather than three separate implementations. Only the client-server link is MCP. The link from the server to PLAXIS 3D is the vendor's own remote-scripting interface, plxscripting over a local TCP port. The MCP server is the adapter between the two. The stack therefore has four layers: the agent platform and its base model (reasoning), the MCP server (bridge), the PLAXIS remote-scripting API that provides the \texttt{g\_i} and \texttt{g\_o} object handles, and the PLAXIS 3D Input and Output engines. Keeping the boundaries distinct allows the top layer to be replaced without touching anything below it.

The server registers 19 tools. Each is a typed schema over one remote-scripting action, so the agent never composes free-form Python. Arguments are validated before they reach PLAXIS. Unknown arguments are rejected rather than silently ignored. Three design decisions each remove a class of silent error observed during development. First, \textbf{no silently wrong defaults on domain-defining arguments}. \texttt{plaxis\_new\_project} requires the four soil-contour extents explicitly. An earlier 10 × 10 m default had produced correct-looking but meaningless models that reached the saved corpus. Second, \textbf{\texttt{extra="forbid"} on every argument model}, so an invented argument name fails loudly. This is the only mechanism that catches a hallucinated key, as opposed to an omitted one. Third, \textbf{a raw command escape hatch} (\texttt{plaxis\_run\_command}, \texttt{plaxis\_list\_commands}). The agent can reach actions the typed surface does not cover, and the server does not need to mirror the whole API.

The 19 tools fall into five groups. \emph{Session:} \texttt{plaxis\_connect} (authenticated connection, binding \texttt{s\_i} and \texttt{g\_i}), \texttt{plaxis\_connection\_status}, \texttt{plaxis\_open\_output} (binds \texttt{g\_o} on a calculated model). \emph{Project:} \texttt{plaxis\_new\_project} (initialises the rectangular soil contour, extents required), \texttt{plaxis\_save\_project}, \texttt{plaxis\_open\_project}. \emph{Model build:} \texttt{plaxis\_add\_borehole}, \texttt{plaxis\_add\_soil\_layer}, \texttt{plaxis\_create\_soil\_material} (\texttt{g\_i.soilmat()} with optional layer assignment), \texttt{plaxis\_set\_property} (optionally phase-dependent), \texttt{plaxis\_goto\_mode}. \emph{Analysis:} \texttt{plaxis\_generate\_mesh} (project refinement recipe at a given coarseness), \texttt{plaxis\_add\_phase}, \texttt{plaxis\_calculate}, \texttt{plaxis\_add\_curve\_point}, \texttt{plaxis\_get\_results}. \emph{Introspection and escape:} \texttt{plaxis\_list\_objects}, \texttt{plaxis\_list\_commands}, \texttt{plaxis\_run\_command}.

Connection uses the official remote-scripting interface. In PLAXIS 3D the path is Expert, then Configure remote scripting server, with a port and a password. An \texttt{mcp.json} file then registers the server with the AI client. The MCP client launches \texttt{plaxis\_mcp\_server.py} and establishes the authenticated TCP connection to the \texttt{g\_i} and \texttt{g\_o} engines.

\section{System Architecture}\label{system-architecture}

\subsection{Four agents, and why the verifier is outside them}\label{four-agents-and-why-the-verifier-is-outside-them}

The modelling task is decomposed across four specialised agents (Fig. 4). They communicate through a shared run-state blackboard and reach PLAXIS 3D only through the MCP server.

\begin{itemize}
\tightlist
\item
  \textbf{O: Orchestrator.} Sequences the agents, owns the blackboard, gates every calculation on the audit, and routes on PASS or FAIL. It holds no PLAXIS tools of its own.
\item
  \textbf{G: Geometry agent.} Builds the project, the Bangkok material sets, the soil profile, the surcharge, the tunnel with its lining and shield, and the staged-construction routine. It does no meshing, groundwater assignment or solving.
\item
  \textbf{C: Calculation agent.} Meshes at calibrated coarseness, assigns the groundwater regime per cluster, defines the ring-advance schedule, self-checks, saves the pre-calculation model, and solves only when the gate opens.
\item
  \textbf{V: Verification agent.} A fresh, read-only, information-barriered agent. It re-derives every modelling convention from the raw specification and audits the built model against it.
\end{itemize}

The verifier's independence is enforced by construction, not by instruction. Its configuration holds raw specification values only. Pressures are held at the reference points the source document quotes them at: face pressure at the axis and grout pressure at the crown. Stratum bottoms are held exactly as stated, and the piezometric level as read. The auditor recomputes every project convention itself. These include the face-axis-to-excavated-crown conversion, the grout reference, the drawdown midpoint band and the layer splits it requires, the phase count, and the contraction ramp. It is forbidden from importing conversion logic from the build scripts, because a shared bug would then be invisible to it. Omitting an optional audit block does not silently pass. The check is skipped, a warning is raised, and the gap is named in the verdict's coverage line.

\subsection{Workflow and pipeline}\label{workflow-and-pipeline}

A specification arrives as human language, a PowerPoint master table, a PDF report or JSON. The intake stage extracts and tags every value and confirms ambiguities. The orchestrator then sequences the agents, and verification enters at two points. The first is the \textbf{pre-mesh checkpoint}, after the geometry agent completes. Here the verifier runs the identity and geometry subset of its checks: model bounds, borehole head, strata, column extent, and the convention splits the groundwater regime requires. A failure here halts the run before any mesh or staging time is spent. The second is the \textbf{full audit gate}, after meshing, groundwater assignment, ring-advance definition and the pre-calculation save. Here the verifier performs the complete independent audit, and only a clean verdict releases the solve. The orchestrator's response to a failing verdict is to halt and surface the findings. It spends no solver time. Automated repair is a natural extension. It is deliberately outside the measured protocol, so no arm can be rescued by retry. If the auditor itself crashes, the run is recorded as an instrument error and re-run rather than scored.

\begin{figure}[!t]
\centering
\includegraphics[width=\columnwidth]{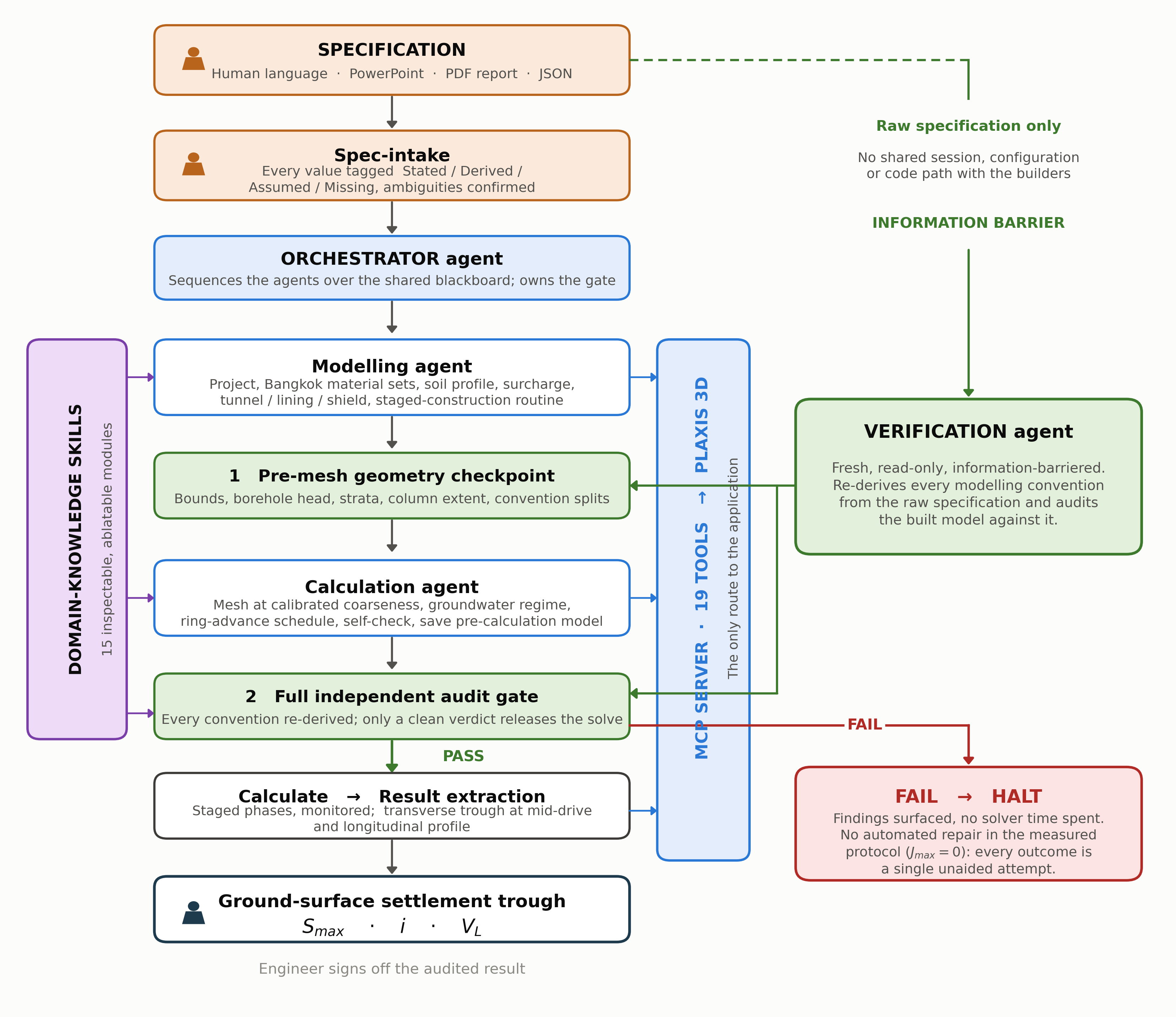}
\caption{Workflow from specification through intake, orchestrated build, information-barriered audit and gated calculation to result extraction.}
\label{fig:4}
\end{figure}

The build follows an 11-stage pipeline (Fig. 5). Every stage drives PLAXIS 3D through the MCP server. Stages 1 to 5 establish the model: units and boundaries, the borehole with its layers and head, the Bangkok hardening-soil materials, the shield with its lining, trajectory and slices, and the surface surcharge where specified. Stages 6 to 8 prepare and run the excavation: the mesh at calibrated coarseness, the groundwater regime, and the ring-by-ring advance. Stage 9 is the full audit. It re-derives stages 1 to 8 from the raw specification and must PASS. Stage 10 calculates the staged phases with monitoring points active. Stage 11 extracts the transverse trough with \(S_{max}\), \(i\) and \(V_L\).

\begin{figure}[!t]
\centering
\includegraphics[width=\columnwidth]{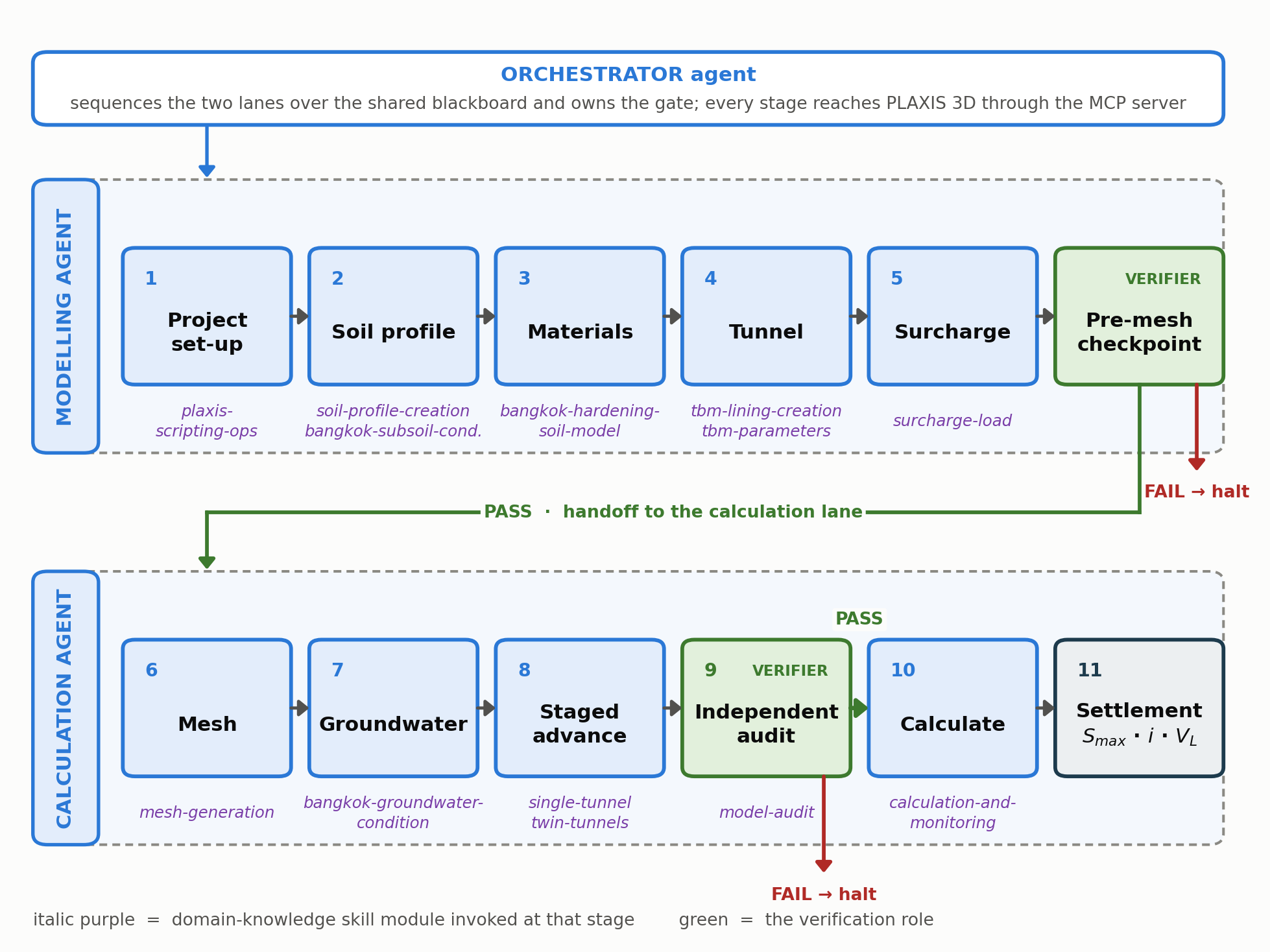}
\caption{Modelling pipeline: 11 stages from confirmed specification to settlement.}
\label{fig:5}
\end{figure}

\subsection{Domain-knowledge skill modules and specification intake}\label{domain-knowledge-skill-modules-and-specification-intake}

Fifteen inspectable, ablatable modules encode the geotechnical and operational knowledge shared by every agent role. Five carry \textbf{Bangkok ground knowledge}:

\begin{itemize}
\tightlist
\item
  the standard 9-layer stratigraphy and its mapping onto model layer stacks;
\item
  per-stratum Hardening Soil parameters (\(\gamma\), \(E_{50}\), \(E_{oed}\), \(E_{ur}\), \(c'\), \(\varphi'\), \(K_0\)) with verbatim soilmat commands;
\item
  the three groundwater regimes and how each is encoded;
\item
  surface surcharge creation and its staging in a pre-drive phase;
\item
  soil-profile creation, covering contour, borehole and layer commands, material assignment and water-table conventions.
\end{itemize}

Four carry \textbf{tunnelling workflow}:

\begin{itemize}
\tightlist
\item
  the master single-tunnel pipeline with its build strategies and verification checklist;
\item
  the twin-tunnel workflow, covering parallel, stacked and offset layouts, two-drive phasing and deep-first sequencing;
\item
  the verified TBM and lining build sequence, covering cross-section, thick lining, trajectory, slicing, phase routine and advance-loop staging;
\item
  authoritative TBM parameters with their sign and reference conventions, covering face pressure at the axis against grout pressure at the crown, Zref, contraction, materials and diameter mapping.
\end{itemize}

Six carry \textbf{operational procedure}:

\begin{itemize}
\tightlist
\item
  the mesh refinement recipe with per-case coarseness calibration and expected element counts;
\item
  specification intake, covering the extraction checklist, provenance tagging, known ambiguity traps and confirm-before-build;
\item
  scripting-session mechanics, covering connection, batching limits, object naming lifecycle, replay skeleton and background execution;
\item
  solver rules with per-phase health monitoring and a failure taxonomy;
\item
  the independent model audit;
\item
  the four-agent decomposition itself, with its shared run-state, audit gate and ablation switches.
\end{itemize}

The intake encodes the conventions behind every specification-reading error observed during development. Every value entering the pipeline is tagged \textbf{Stated} (verbatim from the source), \textbf{Derived} (computed, with the conversion shown), \textbf{Assumed} (a default, flagged for confirmation) or \textbf{Missing} (must be resolved with the user).

\subsection{Formulation}\label{formulation}

Let \(D\) be the raw specification. The intake operator \(\mathcal{I}\) produces the tagged specification

\[\Sigma = \mathcal{I}(D) = \{(k_j, v_j, \tau_j)\}_{j=1}^{m}, \qquad \tau_j \in \{\text{ST}, \text{DE}, \text{AS}, \text{MI}\} \tag{4}\]

Construction is permitted only when no tag remains MI and every AS value has been confirmed. The framework is the tuple \(\mathcal{F} = \langle \mathcal{A}, \mathcal{K}, \mathcal{T}, B \rangle\). Here \(\mathcal{A} = \{O, G, C, V\}\) is the agent set, \(\mathcal{K}\) the 15 skill modules, \(\mathcal{T}\) the MCP tool set and \(B\) the shared blackboard. Each agent \(i\) is an LLM policy \(\pi_i\). It is conditioned on its role prompt, its skill subset \(\mathcal{K}_i \subseteq \mathcal{K}\) and the blackboard. At step \(t\) it emits an action \(a_t = \pi_i(\Sigma, B_t, \mathcal{K}_i; \theta) \in \mathcal{T}_i \cup \{\emptyset\}\), with \(\theta\) the base model. Each executed call advances the model state and the blackboard, \(M_{t+1} = \delta(M_t, a_t)\) and \(B_{t+1} = u(B_t, a_t, r_t)\). The transition \(\delta\) is realised by the MCP server and the scripting API. Rolling out agent \(i\) defines \(\Phi_i\). The build is \(M_{\text{built}} = (\Phi_C \circ \Phi_G)(M_0; \Sigma)\), sequenced by \(O\), which holds no PLAXIS tools (\(\mathcal{T}_O = \emptyset\)).

The verifier is a fresh policy. Its observations are restricted to \(\mathrm{Obs}(V) = \{D, \mathrm{read}(M_{\text{built}})\}\). The tagged specification, the blackboard and the builder transcripts are withheld. \(V\) evaluates \(n_c\) binary checks \(c_k\). The checkpoint uses the geometry subset \(\mathcal{C}_{\text{geo}}\) and the gate uses all of them,

\[g_{\text{geo}} = \bigwedge_{k \in \mathcal{C}_{\text{geo}}} c_k, \qquad g = \bigwedge_{k=1}^{n_c} c_k \tag{5}\]

On the round-1 corpus \(n_c \in [12, 23]\) with mode 18, and \(|\mathcal{C}_{\text{geo}}| = 5\) in the standard configuration. The solver runs only behind the gate: \(M_{\text{solved}} = \Psi(M_{\text{built}})\) if \(g_{\text{geo}} = g = 1\), and HALT otherwise. A repair extension that routes the failing-check set to the responsible agent is defined but not exercised, so \(J_{\max} = 0\) throughout. Every cell is therefore a single unaided attempt, and every reported rate is directly comparable across arms. The reliabilities below are lower bounds on what the deployed system achieves with an engineer or a retry loop in the recovery role. End-to-end reliability factorises as \(P(\text{success}) = P(\text{build}) \cdot P(g = 1 \mid \text{build}) \cdot P(\text{solve} \mid g = 1)\). This separates the failure modes the arms probe.

Every ablation arm modifies a single term. Arm B collapses \(\mathcal{A}\) to one policy, so intake, construction and checking share a context. Arm C removes \(O\). Arms D and E replace \(\Phi_G\) and \(\Phi_C\) with fixed scripted drivers. Arm F removes \(V\) and sets \(g \equiv 1\), with the audit still running post hoc as the measurement instrument. Arm G sets \(\mathcal{K} = \emptyset\). The H-series varies \(\theta\). The P-series varies the host harness against identical \(\mathcal{T}\) and \(\mathcal{K}\). Two arms of an earlier design were dropped when the corpus was rebuilt (Section VII-D). One relaxed the barrier and one bypassed intake. Neither is isolated by a dedicated arm here.

\section{Benchmark, Finite-Element Modelling and Ablation Design}\label{benchmark-finite-element-modelling-and-ablation-design}

\subsection{Problem benchmark}\label{problem-benchmark}

The scored benchmark comprises 12 single-tunnel problems across four capability tiers (Fig. 6). The tiers are nested so that each isolates one axis of demand against an otherwise fixed case. Tier A varies tunnel depth. Tier B varies only the TBM operating parameters on the A2 ground. Tier C varies only the tunnel and machine geometry on the A4 ground. Tier F varies only the surface and groundwater boundary conditions at the A3 axis elevation on profile P2.

\begin{figure*}[!t]
\centering
\includegraphics[width=\textwidth,height=0.48\textheight,keepaspectratio]{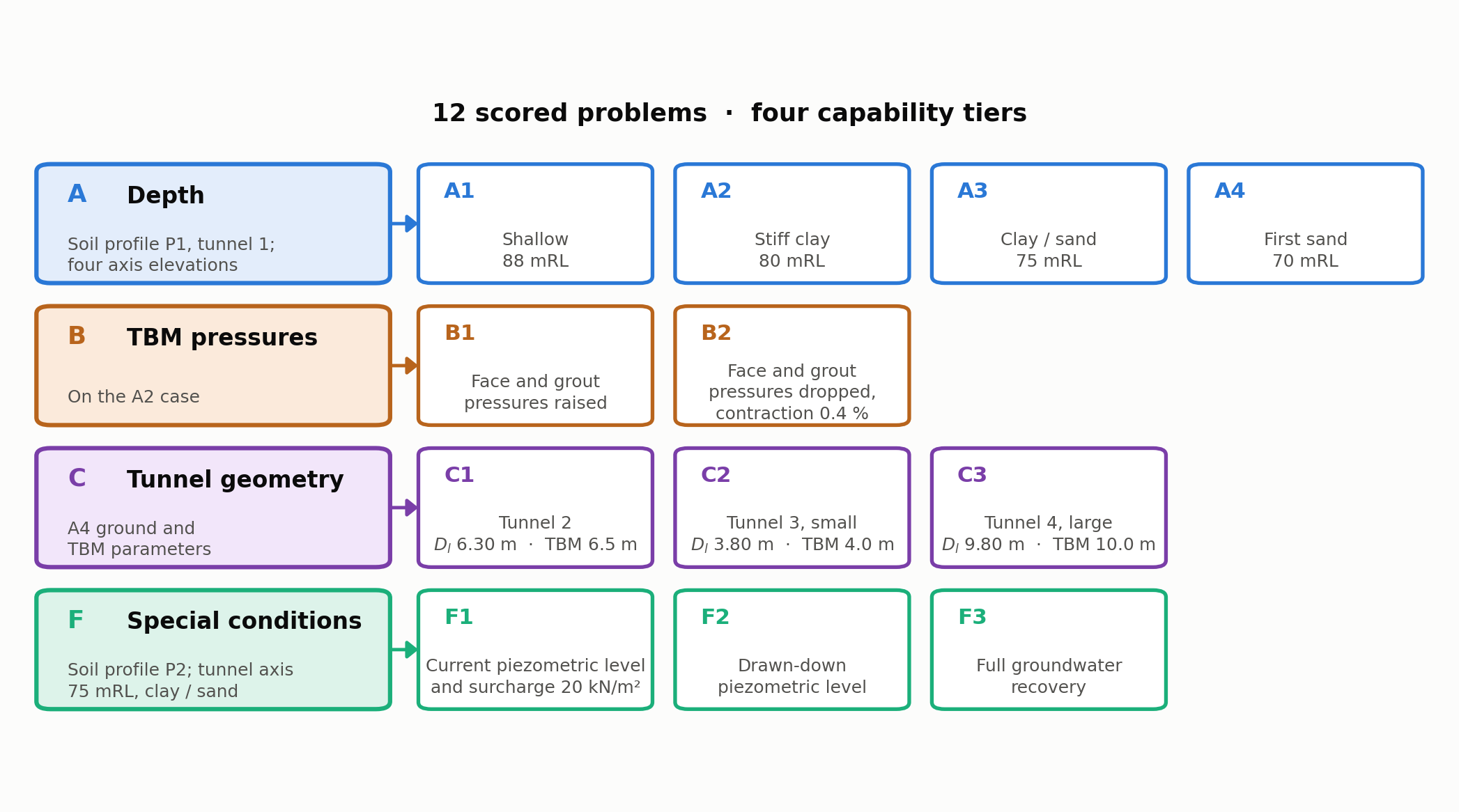}
\caption{Problem benchmark: capability tiers A, B, C and F (12 single-tunnel problems).}
\label{fig:6}
\end{figure*}

Two idealised Bangkok profiles are used (Fig. 7). Elevations are in mRL, with the surface at 100 mRL. Three groundwater regimes are considered: the current condition (piezometric level 89 mRL, drawdown from deep-well pumping), deep drawdown (80 mRL) and hydrostatic (100 mRL, long-term recovery). Problem F1 adds a uniform 20 kN/m² surcharge on profile P2. Soil layers use the Hardening Soil model with Bangkok parameters after Surarak et al.~(2012) and Likitlersuang et al.~(2014). Each layer carries its unit weight, strength parameters (\(c'\), \(\varphi'\)), stiffness moduli and drainage type. Made ground and sand are drained. The clays are undrained.

\begin{figure*}[!t]
\centering
\includegraphics[width=\textwidth,height=0.48\textheight,keepaspectratio]{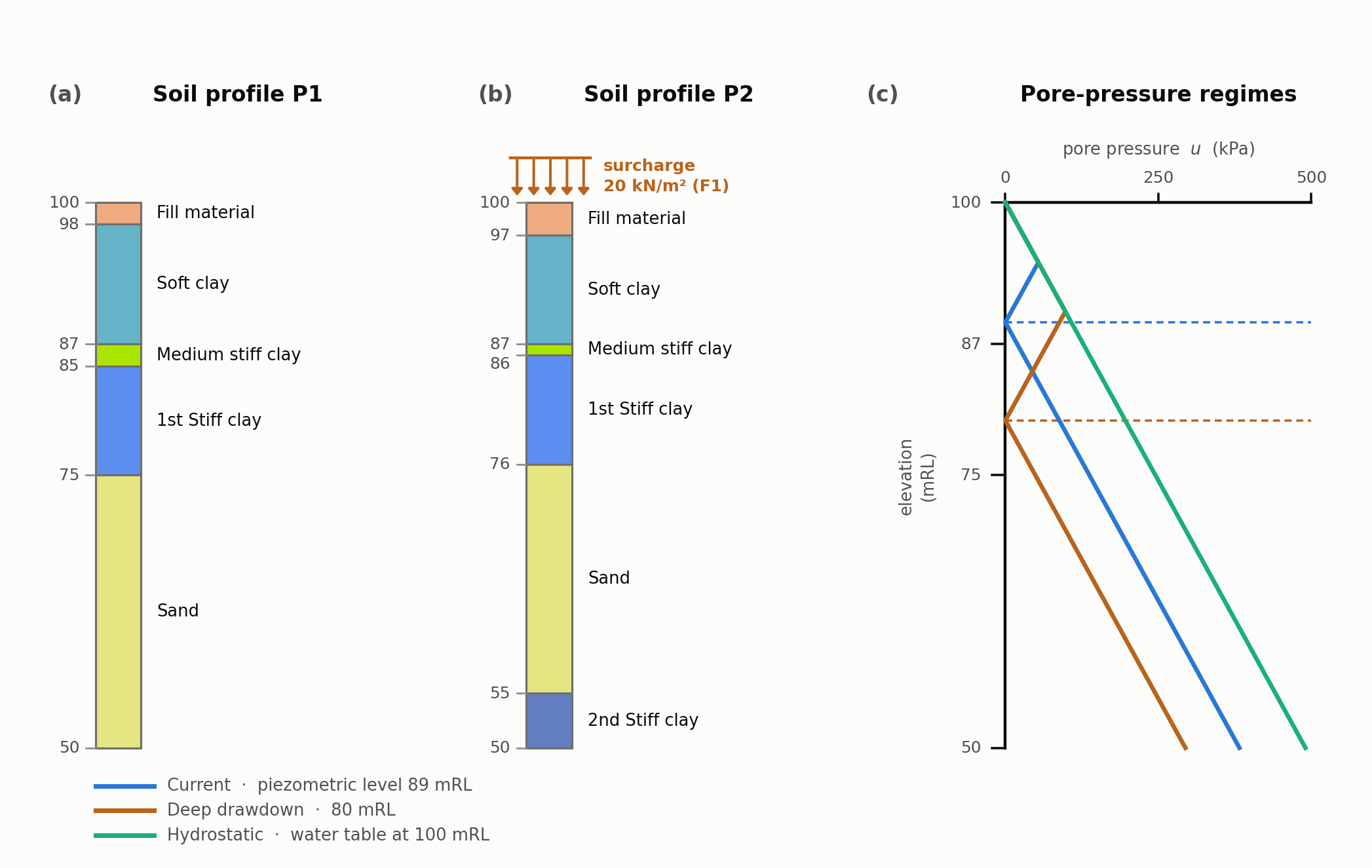}
\caption{Benchmark ground conditions: (a) profile P1, (b) profile P2 with the 20 kN/m² surcharge of F1, (c) the three pore-pressure regimes.}
\label{fig:7}
\end{figure*}

Four tunnel and TBM combinations are defined (Fig. 8, Table II), with 0.5 \% shield conicity throughout.

\begin{table*}[!t]
\caption{Tunnel and TBM geometry.}
\label{tab:2}
\centering\footnotesize
\setlength{\tabcolsep}{4pt}\renewcommand{\arraystretch}{1.15}
\begin{tabularx}{\textwidth}{Ylllllll}
\toprule
Tunnel & $D_l$ (m) & $D_i$ (m) & Lining $t$ (m) & Ring $L_r$ (m) & Shield $D_s$ (m) & Shield $L_s$ (m) & $C_{ref}$ (\%) \\
\midrule
1: MRT typical detail 1 & 6.30 & 5.70 & 0.30 & 1.4 & 6.6 & 9.8 & 0.5 \\
2: MRT typical detail 2 & 6.30 & 5.70 & 0.30 & 1.2 & 6.5 & 9.6 & 0.5 \\
3: Small & 3.80 & 3.44 & 0.18 & 1.0 & 4.0 & 7.0 & 0.5 \\
4: Large & 9.80 & 9.10 & 0.35 & 1.5 & 10.0 & 12.0 & 0.5 \\
\bottomrule
\end{tabularx}
\end{table*}

\begin{figure*}[!t]
\centering
\includegraphics[width=\textwidth,height=0.48\textheight,keepaspectratio]{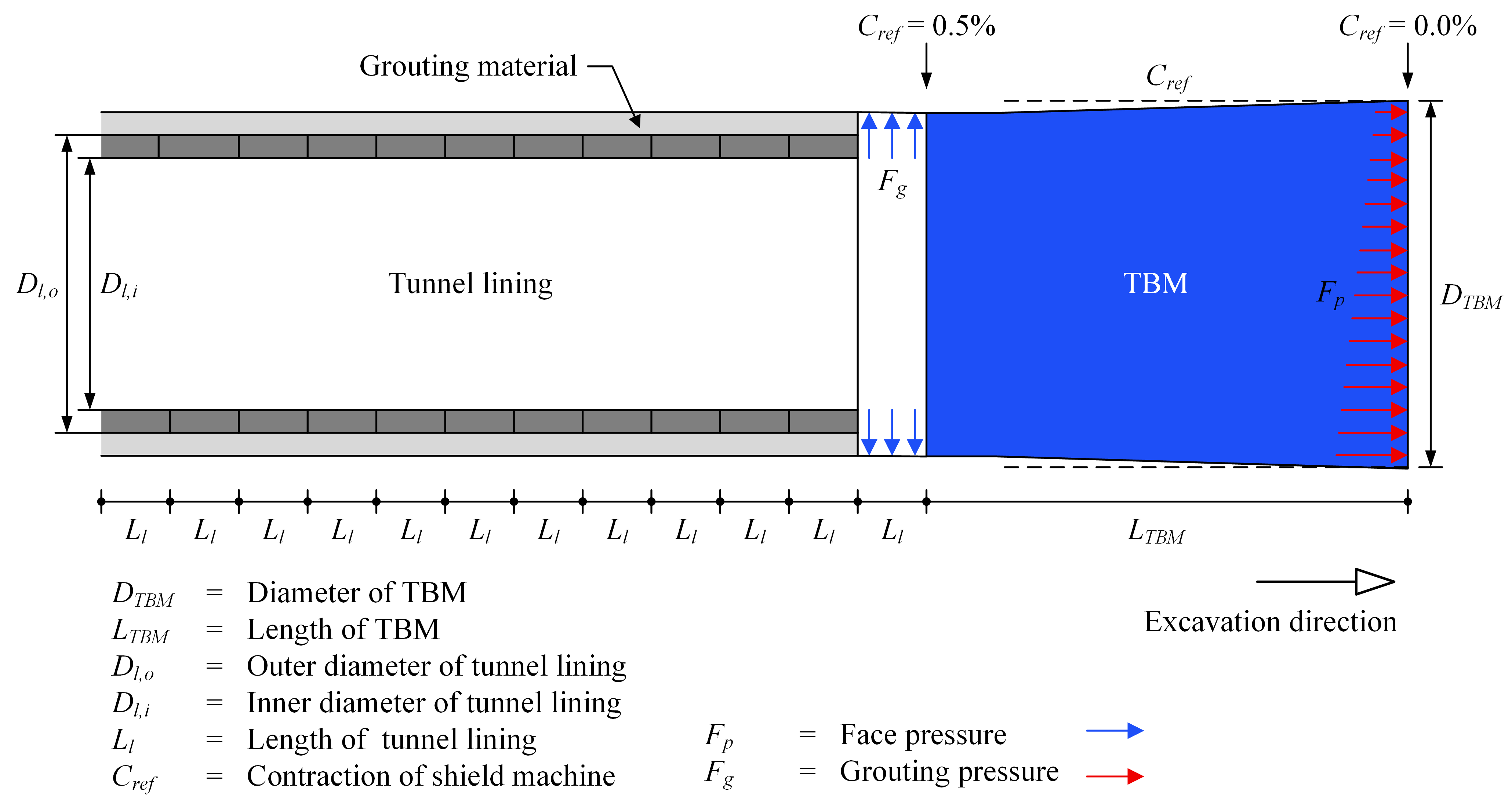}
\caption{TBM and tunnel lining.}
\label{fig:8}
\end{figure*}

Two operating pressures are specified per problem. Face pressure is quoted at the tunnel axis and applied as a trapezoidal distribution increasing with depth. Grouting pressure is quoted at the crown and applied over the tail void. The reference point matters as much as the value. The builder must convert each stated pressure to the reference the software expects. That conversion is one of the checks the auditor re-derives independently. Both pressures rise with axis depth across tier A, from 160/190 kN/m² at 88 mRL to 280/330 kN/m² at 70 mRL. Tier B varies them on the fixed A2 ground (B1 230/270, B2 180/200). Tier C uses 280/330 throughout. Tier F uses 230/280 (F1, F2) and 300/350 (F3).

\subsection{Finite-element modelling}\label{finite-element-modelling}

Each problem is built as a single 3D model (Fig. 9). The mesh is fine around the tunnel and coarsens towards the boundaries. The domain measures 100 m along the drive, about 80 m across and 50 m deep. Along the drive this satisfies the clearance of not less than \(4D_s\) ahead of and behind the excavation face recommended for Bangkok tunnelling (Phutthananon et al., 2023). The transverse trough is extracted at a monitoring section about 40 m along the drive, clear of both ends. The lateral sides are restrained horizontally and free vertically. The bottom is fully fixed and the surface is unrestrained. All boundaries except the top surface are impermeable.

\begin{figure}[!t]
\centering
\includegraphics[width=\columnwidth]{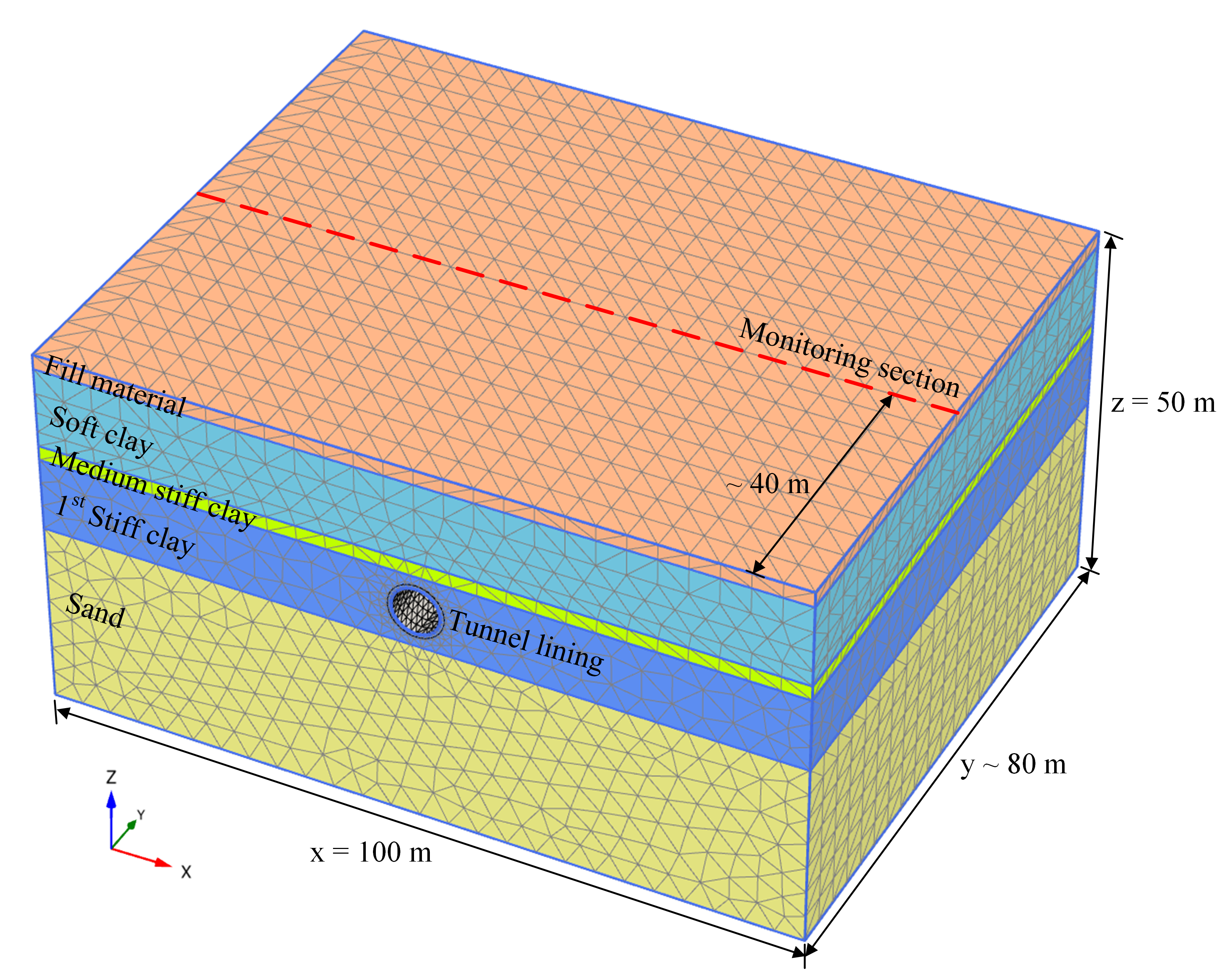}
\caption{Three-dimensional finite-element model: mesh, stratigraphy, tunnel lining and monitoring section.}
\label{fig:9}
\end{figure}

The staged EPB-shield procedure follows Phutthananon et al.~(2023) (Fig. 10). The shield advances ring by ring. A face pressure acts at the excavation face, contraction is applied along the conical shield, grout pressure acts at the tail, and the segmental lining is activated behind it. The conventions adopted are theirs: one advance step per lining ring, a trapezoidal face pressure increasing with depth, a calibrated contraction ratio on the shield shell, and a two-phase tail-void treatment in which liquid grout pressure is followed by hardened grout with the lining activated.

\begin{figure}[!t]
\centering
\includegraphics[width=\columnwidth]{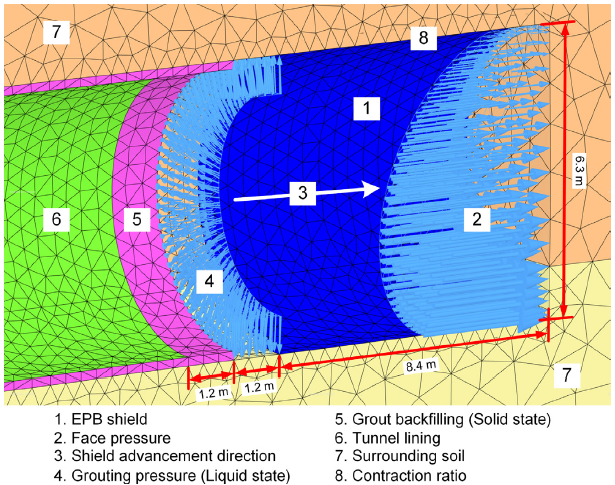}
\caption{Components of the EPB shield advancement analysis. Reproduced from Phutthananon et al. (2023).}
\label{fig:10}
\end{figure}

\subsection{Ablation design and protocol}\label{ablation-design-and-protocol}

The ground truth for every problem is \textbf{arm A}. The same 12 problems were built by hand in the graphical interface by an experienced engineer, with no agent and no MCP. It is built once, because a manual reference does not replicate, and it is the only arm exempt from the replication design. Fifteen arms are defined (Table III). They cover the manual baseline, four architecture ablations, one verification ablation, the knowledge ablation, four capability rungs and four portability arms. The fourteen automated arms drive the same MCP server, builder scripts and skill modules, except where the arm's definition removes them. Ablation is applied through an orchestrator switch (\texttt{-\/-no-orchestrator}, \texttt{-\/-no-geometry}, \texttt{-\/-no-calculation}, \texttt{-\/-no-gate}). For the knowledge arm, the skill directory is renamed out of the agents' reach for the run. Capability arms differ in exactly one respect, the model pinned in the four role definitions. Portability arms differ in exactly one respect, the host harness.

\begin{table*}[!t]
\caption{Ablation arms. O, G, C and V are the orchestrator, geometry, calculation and verification agents.}
\label{tab:3}
\centering\footnotesize
\setlength{\tabcolsep}{4pt}\renewcommand{\arraystretch}{1.15}
\begin{tabularx}{\textwidth}{llYY}
\toprule
Arm & Category & What is removed or changed & Harness / model \\
\midrule
A & Baseline & Manual workflow in the GUI: no agent, no MCP & human expert \\
B & Architecture & Single agent; whole pipeline fused in one context, self-audit only & Claude Code / Opus 5 \\
C & Architecture & O removed: fixed linear sequence, no checkpoint, audit non-gating & Claude Code / Opus 5 \\
D & Architecture & G removed: construction via a fixed scripted driver & Claude Code / Opus 5 \\
E & Architecture & C removed: mesh, water, phases and save via a fixed scripted driver & Claude Code / Opus 5 \\
F & Verification & V removed: no checkpoint, no gate; audit runs post hoc as instrument only & Claude Code / Opus 5 \\
G & Knowledge & All 15 skill modules withheld; general reasoning only & Claude Code / Opus 5 \\
H1–H4 & Capability & Full system, base model varied & Claude Code / Fable 5, Opus 5 (reference), Sonnet 5, Haiku 4.5 \\
P1–P3 & Portability & Full system, different harness, same MCP server & GPT-Codex / GPT-5.6 Sol, Terra, Luna \\
P4 & Portability & Full system, third vendor (defined, not executed) & Kimi Code / kimi-k3 \\
\bottomrule
\end{tabularx}
\end{table*}

Crossing 15 arms with 12 problems yields 180 cases. This is round 1. Two further rounds of the 14 automated arms are in progress, giving 516 cells over three rounds. The results below cover the ten Claude arms and three GPT-Codex arms over all 12 problems. Two references do different work. \textbf{Arm H2 is the concordance reference.} It is an output of the system under test. The mesh fingerprint and between-arm agreement are measured against it. \textbf{Arm A is the accuracy reference.} It is external to the system.

Each cell is a single unaided attempt: one fresh specification extraction, one build, no automated repair and no operator intervention. The intake's confirmation step is therefore disabled for the campaign. Assumed values are recorded with their tags and defaulted rather than referred to the engineer. Models are built and audited on one machine and transferred to a faster machine for solving. Both run the same PLAXIS version, because the mesher changes between versions. Build time excludes solver time. A watchdog terminates any build exceeding 75 min of wall-clock modelling time.

Arms are scored on four quantities. \textbf{Build reliability} \(R = N_{built}/N_{att}\) is the share of attempts that produce a valid saved model. \textbf{Construction accuracy} is the share that return a clean independent audit. \textbf{Settlement error} is the mean relative deviation in \(S_{max}\) and \(i\) from arm A. \textbf{Build time} is wall-clock modelling time. Several arms sit at \(R = 0\) or \(R = 1\), where the normal approximation is degenerate, so interval estimates are Wilson score intervals. At \(N = 12\) these give {[}0, 24.3 \%{]} for a total halt and {[}75.8 \%, 100 \%{]} for a clean sweep. The intervals are wide enough that only large effects are resolvable. Round 1 provides n = 1 per cell, so the intervals quantify sampling uncertainty across problems but not run-to-run variance. Rounds 2 and 3 re-run the full pipeline, including fresh specification extraction. Replaying stored configurations would drive apparent variance to near zero. It would also miss specification reading, which is where the arms actually differ.

A fifth quantity carries more signal than any of these. The \textbf{mesh fingerprint} of a cell is its element count compared against the concordance reference. The mesh follows from the geometry, stratigraphy, tunnel, surcharge and refinement recipe. Its element count is therefore a sensitive scalar summary of the whole construction. Two models that agree to the element give strong evidence that the upstream geometric decisions agreed. The test is necessary rather than sufficient. It is insensitive to anything that does not change the mesh, such as material parameters, pressures or phase ordering. It is available for every built cell without solving. That makes it the only accuracy-like measure that does not require a completed solve.

\section{Results}\label{results}

\subsection{Representative case H2\_A2}\label{representative-case-h2_a2}

Case H2\_A2 illustrates the end-to-end product. It is the full system with Opus 5 on the tier-A stiff-clay single tunnel. The axis is at 80 mRL in the 1st stiff clay of profile P1, 20 m below the surface. The case is also the concordance reference for problem A2. Every other arm that built this problem produced an identical mesh. Six of those have been solved and return an identical trough.

The model comprises 96 978 elements. Figure 11 collects the Output views. Settlement concentrates in a trough above the tunnel and decays towards the boundaries. The vertical-displacement views along the drive confirm that the largest movements track the advancing shield. Figure 12 sets the extracted trough against the model the expert built by hand and against the empirical prediction a Bangkok engineer would make without running anything. The manual and automated troughs are indistinguishable over the full 100 m section. Referenced to the far field, they reach 12.87 mm and 12.83 mm at the centreline and share a trough width of 9.80 m. Both track the empirical Gaussian closely out to about two trough widths and then part from it. They close with a small heave lobe of about 0.5 mm outside 22 m, which the monotonic Gaussian cannot represent. The empirical curve describes the trough well between the inflection points and progressively worse in the far shoulders.

\begin{figure*}[!t]
\centering
\includegraphics[width=\textwidth,height=0.48\textheight,keepaspectratio]{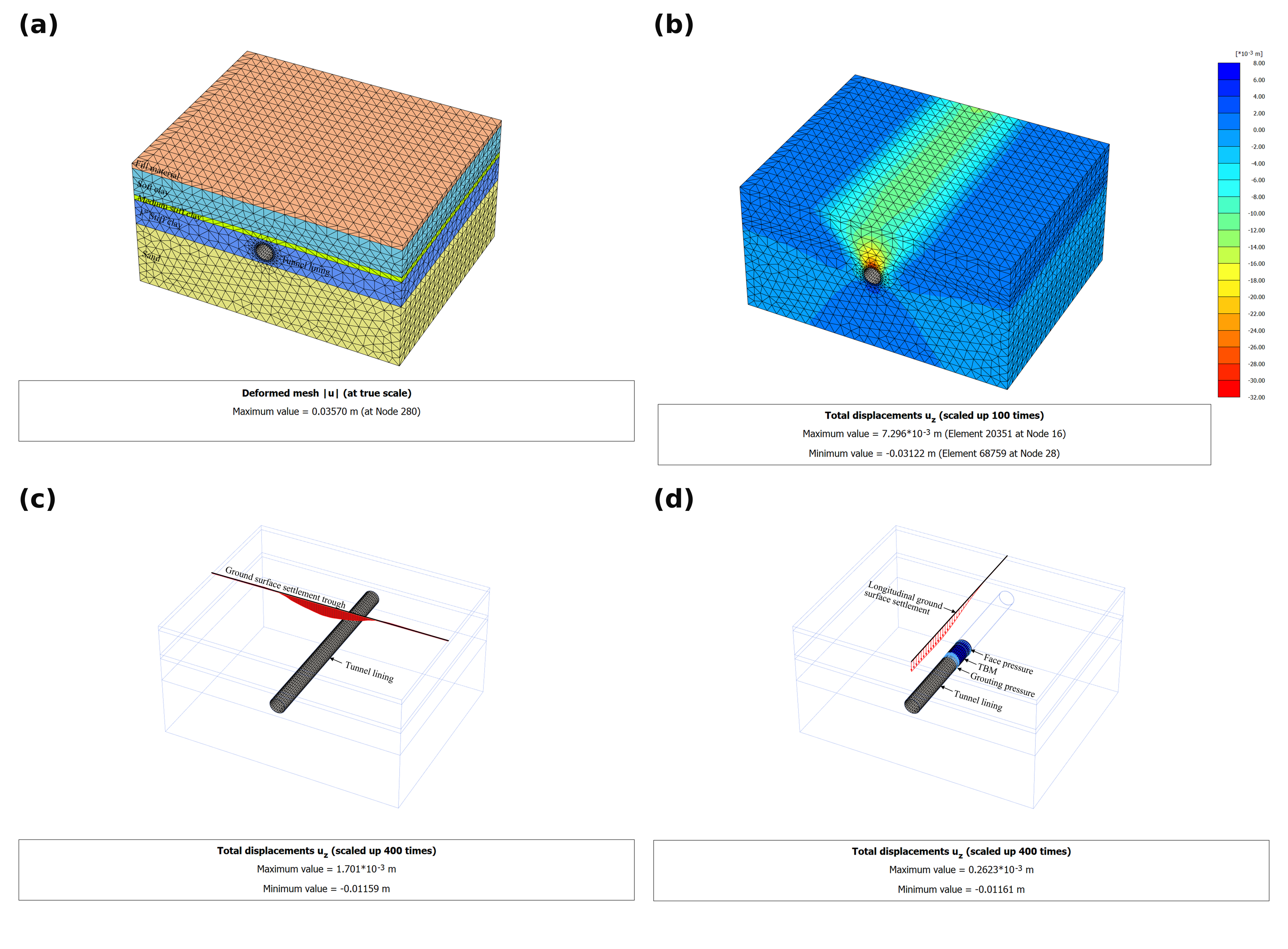}
\caption{Case H2\_A2 in PLAXIS 3D Output: (a) deformed mesh at true scale; (b) vertical displacement $u_z$; (c) transverse settlement trough; (d) longitudinal settlement with shield, face pressure and tail grouting pressure.}
\label{fig:11}
\end{figure*}

\begin{figure}[!t]
\centering
\includegraphics[width=\columnwidth]{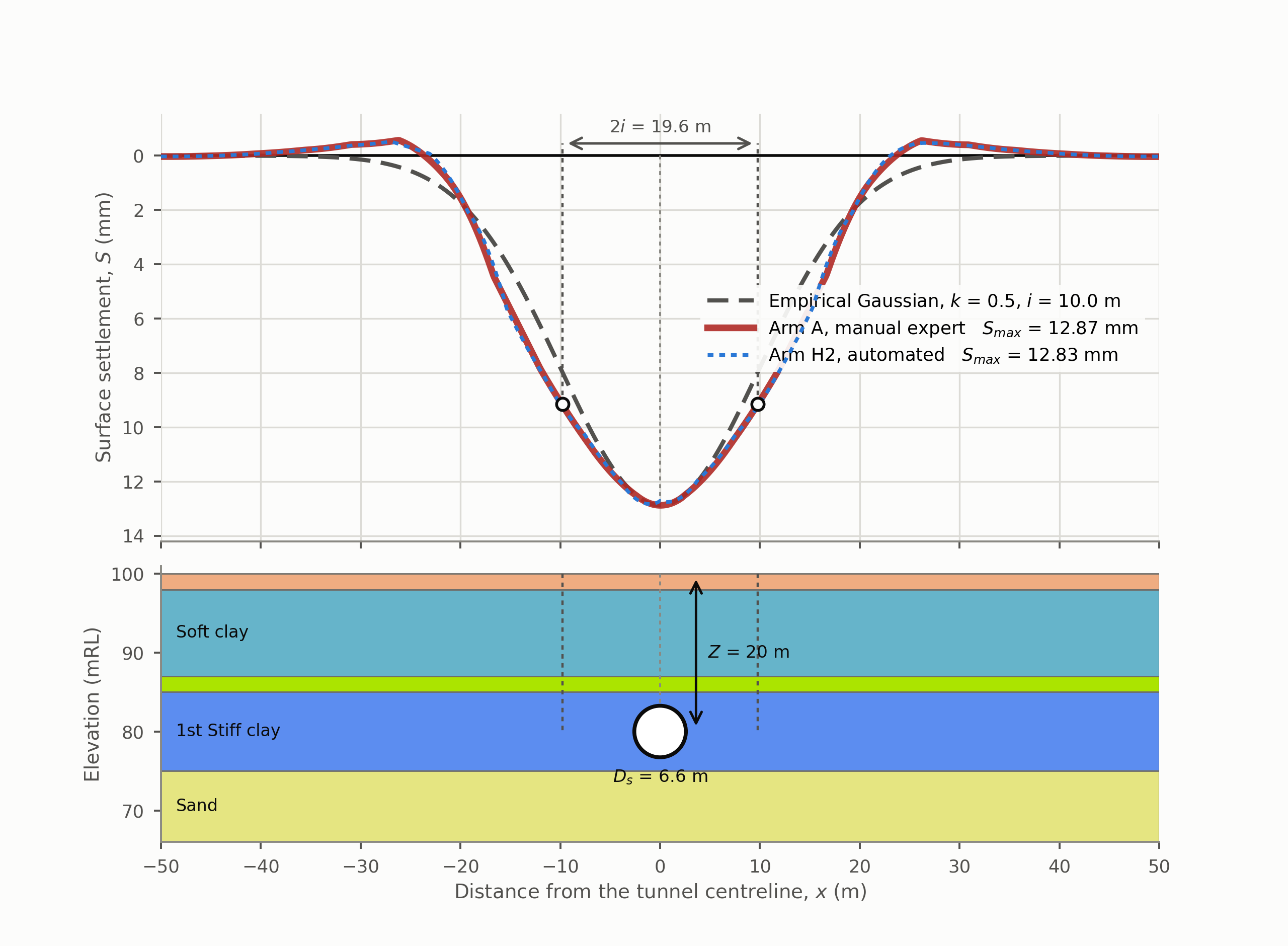}
\caption{Transverse trough at the monitoring section of problem A2, referenced to the far field. The manual expert model (arm A) and the automated model (arm H2) are drawn against the empirical Gaussian of Eq. (1) with $k = 0.5$. The curves are scaled to the same maximum, so the comparison is of shape rather than magnitude. Open circles mark the inflection points at $x = \pm i$.}
\label{fig:12}
\end{figure}

Fitting Eq. (1) to the extracted profile gives \(S_{max}\) = 11.64 mm and \(i\) = 9.80 m (\(R^2\) = 0.930 over \(x = \pm 50\) m). With the axis 20 m below the surface, \(k = i/Z = 0.49\). The trough volume is referenced to the excavated diameter \(D_s\) = 6.6 m. The lining outer diameter \(D_l\) is 6.30 m, and the two must not be interchanged. This gives \(V_s\) = 0.286 m²/m and \(V_L\) = 0.84 \%. All three values fall inside the ranges reported for Bangkok MRT bored tunnelling. \(k\) = 0.49 is close to the \(i = 0.5Z\) of O'Reilly and New (1982) and mid-band for Mair and Taylor (1997). \(V_L\) = 0.84 \% sits inside the 0.5 to 1.0 \% band that Duangsano et al.~(2023) report for most Bangkok drives. Table IV lists the corresponding fits for all 12 benchmark problems.

\begin{table*}[!t]
\caption{Gaussian trough fits for the full-system model of each benchmark problem (arm H2). Troughs are extracted at mid-drive over $x = \pm 50$ m. $Z$ is the axis depth below the surface. $V_L$ is referenced to the shield diameter.}
\label{tab:4}
\centering\footnotesize
\setlength{\tabcolsep}{4pt}\renewcommand{\arraystretch}{1.15}
\begin{tabularx}{\textwidth}{Yllllll}
\toprule
Problem & $Z$ (m) & $S_{max}$ (mm) & $i$ (m) & $k$ & $V_L$ (\%) & $R^2$ \\
\midrule
A1: shallow, axis 88 mRL & 12 & 13.16 & 5.25 & 0.438 & 0.506 & 0.880 \\
A2: stiff clay, axis 80 mRL & 20 & 11.64 & 9.80 & 0.490 & 0.836 & 0.930 \\
A3: clay/sand interface, axis 75 mRL & 25 & 9.84 & 12.55 & 0.502 & 0.905 & 0.960 \\
A4: deep sand, axis 70 mRL & 30 & 7.54 & 17.70 & 0.590 & 0.978 & 0.992 \\
B1: increased face/grout pressure & 20 & 10.72 & 9.45 & 0.473 & 0.742 & 0.919 \\
B2: contraction 0.4 \% & 20 & 11.32 & 9.85 & 0.493 & 0.817 & 0.930 \\
C1: tunnel 2 & 30 & 7.43 & 17.65 & 0.588 & 0.991 & 0.992 \\
C2: small tunnel 3 & 30 & 3.53 & 17.55 & 0.585 & 1.235 & 0.993 \\
C3: large tunnel 4 & 30 & 14.85 & 18.15 & 0.605 & 0.860 & 0.994 \\
F1: surcharge 20 kN/m² & 25 & 10.06 & 12.40 & 0.496 & 0.914 & 0.976 \\
F2: drawn-down piezometric level & 25 & 9.07 & 12.40 & 0.496 & 0.824 & 0.973 \\
F3: full groundwater recovery & 25 & 8.79 & 12.35 & 0.494 & 0.796 & 0.981 \\
\bottomrule
\end{tabularx}
\end{table*}

The tier-A sequence reproduces the expected physical trend without having been given it. As the axis deepens from 12 m to 30 m, the maximum settlement falls monotonically from 13.2 mm to 7.5 mm. The trough widens from \(i\) = 5.25 m to 17.70 m. The parameter \(k\) rises from 0.44 towards 0.59, consistent with the transition from clay into sand. Volume loss rises from 0.51 \% to 0.98 \% and stays below 1 \% throughout. The Gaussian fit quality improves with depth, from \(R^2\) = 0.88 to 0.99. This reflects the known departure from the Gaussian form for shallow tunnels.

\subsection{Ablation results}\label{ablation-results}

Fig. 13 shows how the arms separate on build reliability, the audit, the mesh fingerprint and build time. No single axis orders them the same way, so the dimensions are reported separately rather than combined. Figure 14 plots build times per cell against the manual baseline. Arm A averages 24.6 min per model (SD 4.8), against 3.4 to 8.1 min for the automated arms.

\begin{figure*}[!t]
\centering
\includegraphics[width=\textwidth,height=0.48\textheight,keepaspectratio]{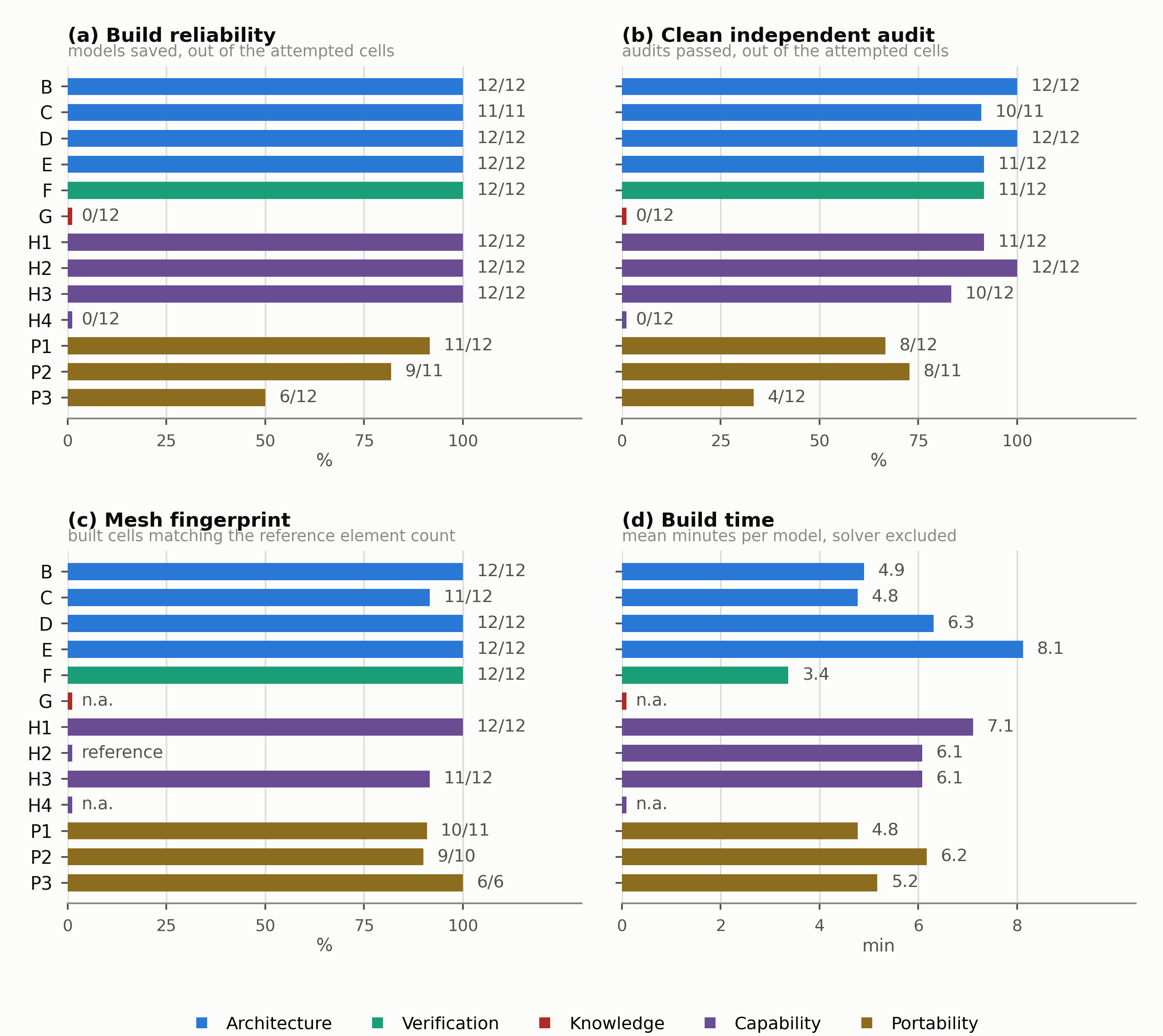}
\caption{The scoring dimensions, with arms held in the same order down every panel. Build reliability separates only the arms that build nothing and the portability arms. The audit separates a different set. The mesh fingerprint separates almost nothing. Build time separates arms that are indistinguishable on every other axis.}
\label{fig:13}
\end{figure*}

\begin{figure*}[!t]
\centering
\includegraphics[width=\textwidth,height=0.48\textheight,keepaspectratio]{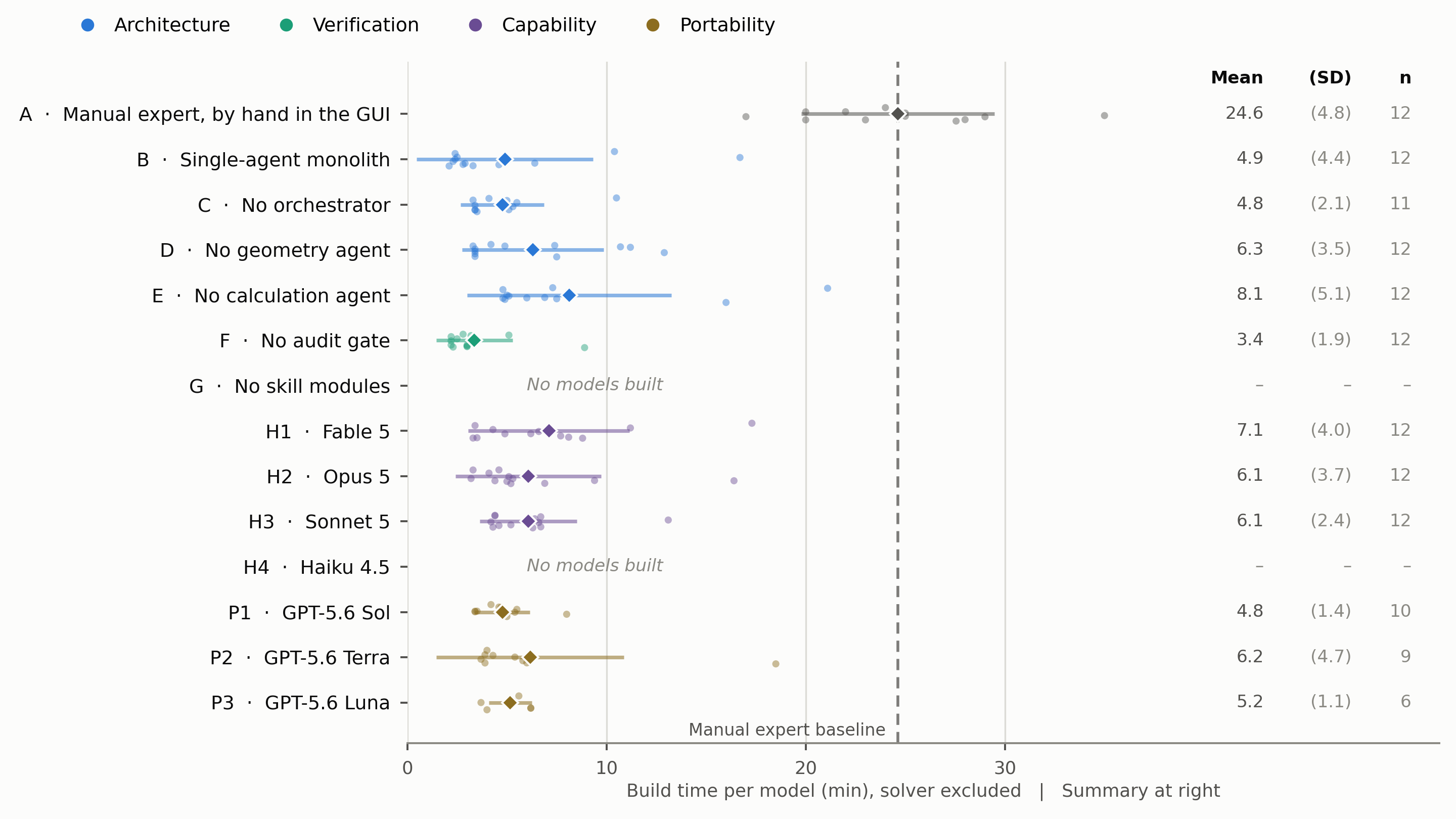}
\caption{Build time per arm against the manual baseline. Each mark is one problem. The diamond is the arm mean, the bar one standard deviation, and the dashed rule the mean of arm A. Arms G and H4 built no models.}
\label{fig:14}
\end{figure*}

Token cost per model (Fig. 15) is far steadier than build time. Within an arm, the dearest case exceeds the cheapest by a factor of 1.3 to 1.7, against 1.7 to 8.0 for build time. A model costs the full system 0.49 M billable tokens on average (SD 0.06 M). The single-agent monolith is the most expensive arm at 0.68 M, because its fused context carries the whole pipeline's material in every turn. Decomposed arms cluster between 0.49 and 0.59 M. The architectural point is in panel (b). The independent auditor accounts for 55 to 60 \% of the tokens in every decomposed arm. Arm F removes the gate but still runs the audit, and it saves nothing (0.53 M against 0.49 M). What the gate removes is the halt, and a halt costs no tokens. Only the monolith spends a minority of its tokens on verification (46 \%). It does so by folding the self-check into the context that built the model. That is precisely the dependence the barrier exists to prevent.

\begin{figure*}[!t]
\centering
\includegraphics[width=\textwidth,height=0.48\textheight,keepaspectratio]{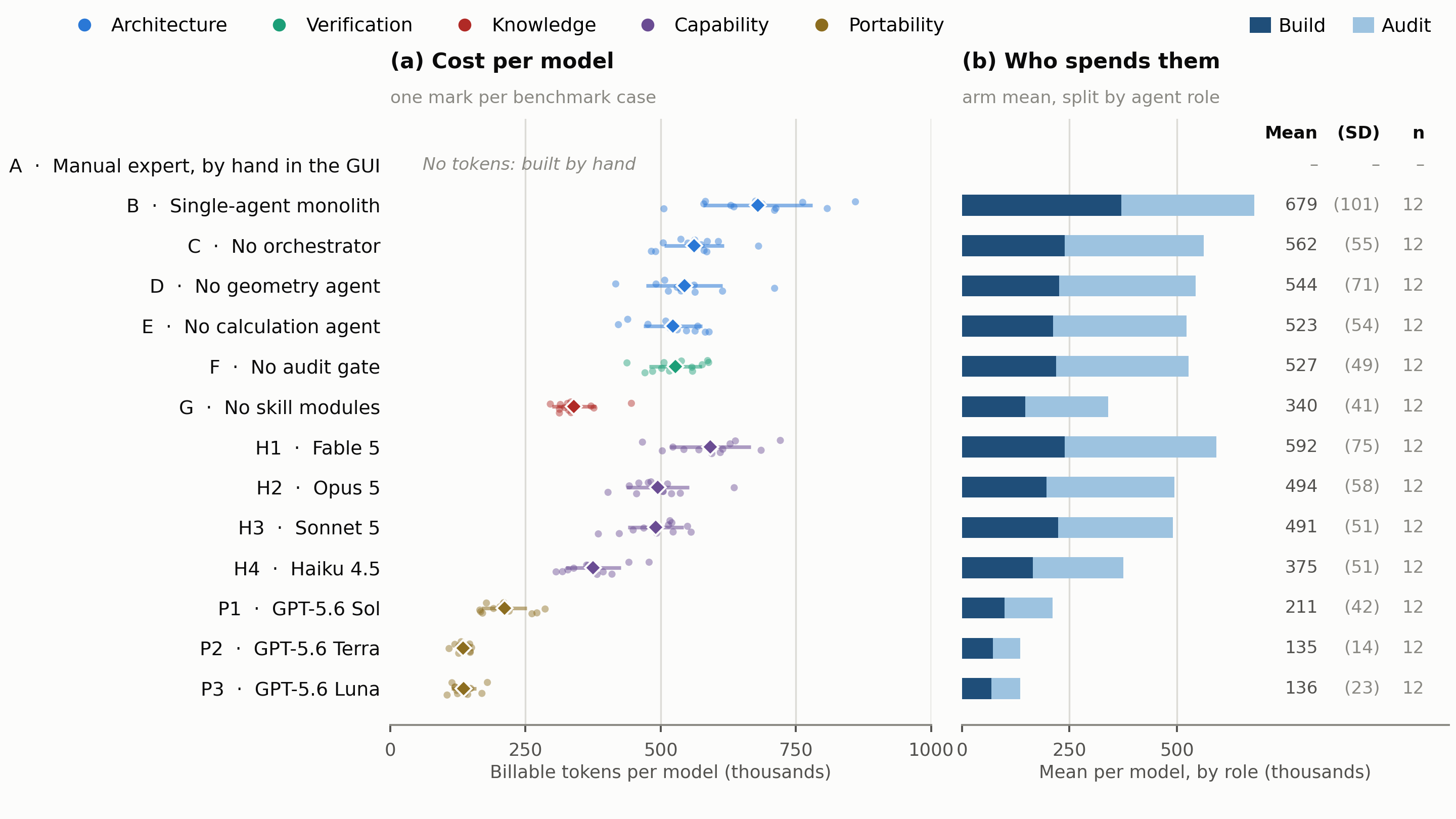}
\caption{Token cost per model built. (a) Billable tokens per case. (b) Arm means split between build and audit roles. Billable tokens are fresh input plus cache creation plus output. Cache reads are excluded. Only the final attempt of each case is counted. The GPT-Codex figures come from that vendor's run logs and are comparable in magnitude but not in accounting.}
\label{fig:15}
\end{figure*}

Table V summarises the key quantitative outcomes across every executed arm.

\begin{table*}[!t]
\caption{Per-arm performance on the 12 benchmark problems (round 1, n = 1 per cell). Reliability intervals are Wilson 95 \% intervals. Build time excludes solver time. No audit instrument runs on a manual build.}
\label{tab:5}
\centering\footnotesize
\setlength{\tabcolsep}{4pt}\renewcommand{\arraystretch}{1.15}
\begin{tabularx}{\textwidth}{lYYlYlY}
\toprule
Arm & Configuration & Harness / model & Built & Build rate [95 \% CI] & Audit PASS & Build time, min (SD) \\
\midrule
A & Manual expert baseline & By hand in the GUI & 12/12 & 100 \% [75.8, 100] & – & 24.6 (4.8) \\
B & Single-agent monolith & Claude Code / Opus 5 & 12/12 & 100 \% [75.8, 100] & 12/12 & 4.9 (4.4) \\
C & No orchestrator & Claude Code / Opus 5 & 11/11 & 100 \% [74.1, 100] & 10/11 & 4.8 (2.1) \\
D & No geometry agent & Claude Code / Opus 5 & 12/12 & 100 \% [75.8, 100] & 12/12 & 6.3 (3.5) \\
E & No calculation agent & Claude Code / Opus 5 & 12/12 & 100 \% [75.8, 100] & 11/12 & 8.1 (5.1) \\
F & No audit gate & Claude Code / Opus 5 & 12/12 & 100 \% [75.8, 100] & 11/12 & 3.4 (1.9) \\
H1 & Full system & Claude Code / Fable 5 & 12/12 & 100 \% [75.8, 100] & 11/12 & 7.1 (4.0) \\
H2 & Full system (reference) & Claude Code / Opus 5 & 12/12 & 100 \% [75.8, 100] & 12/12 & 6.1 (3.7) \\
H3 & Full system & Claude Code / Sonnet 5 & 12/12 & 100 \% [75.8, 100] & 10/12 & 6.1 (2.4) \\
P1 & Portability & GPT-Codex / GPT-5.6 Sol & 11/12 & 91.7 \% [64.6, 98.5] & 8/12 & 4.8 (1.4) \\
P2 & Portability & GPT-Codex / GPT-5.6 Terra & 9/11 & 81.8 \% [52.3, 94.9] & 8/11 & 6.2 (4.7) \\
P3 & Portability & GPT-Codex / GPT-5.6 Luna & 6/12 & 50.0 \% [25.4, 74.6] & 4/12 & 5.2 (1.1) \\
G & No skill modules & Claude Code / Opus 5 & 0/12 & 0 \% [0, 24.3] & 0/12 & n.a. \\
H4 & Full system & Claude Code / Haiku 4.5 & 0/12 & 0 \% [0, 24.3] & 0/12 & n.a. \\
\bottomrule
\end{tabularx}
\end{table*}

Four findings stand out.

\textbf{1. Domain skills are decisive, and they fail at the conventions, not the code.} Arm G has the complete architecture, the full tool set and the strongest base model, but all 15 modules withheld. It produced no model at all. This is the largest single effect in the study, and the failure is legible. In 11 of 12 cells the run died at material assignment (\texttt{set\ Soillayer\_N.Soil.Material\ …}). The agent could drive PLAXIS through the tools. It could not produce the Bangkok Hardening-Soil material sets or the layer-to-material mapping they attach to. The twelfth cell died at the pre-mesh checkpoint on missing convention splits. The drawdown regime requires the profile to be split at the midpoint band, here 89.0 and 94.5 mRL, before the piezometric interpolation can be assigned. Time to failure is bimodal. Eleven cells died within 0.2 min. Cell G\_A1 ran 127.6 min of plausible-looking work before the checkpoint caught the missing splits. That is the most expensive failure mode to detect without an independent checker.

\textbf{2. Base-model capability is a threshold, not a gradient.} With architecture and skills fixed, Fable 5, Opus 5 and Sonnet 5 each build 12 of 12. Haiku 4.5 halts on all 12. There is no intermediate rung in this corpus. The Haiku failures are specification-reading failures rather than modelling failures. They occurred at the checkpoint within 0.3 min on average. Four cells set the model bounds wrongly, and one of these adopted the 1.4 m ring length as the domain extent. Three cells missed the drawdown splits. One assigned a fill unit weight of 18.5 kN/m³ where the specification states 17. Four crashed before a verdict could be recorded. Arm H4 ran under the pre-change harness configuration of Section VII-D, so its total halt is provisional. The ordering is not a restatement of general agentic capability. Arms G and H2 run the same base model and span 0 \% to 100 \% depending only on whether the knowledge layer is present.

\textbf{3. Once the knowledge layer is present, neither the decomposition nor the gate is what produces a correct model.} This negative result cuts against the study's own prior. Removing the orchestrator (C), the geometry agent (D), the calculation agent (E) or the entire verification role (F) leaves build reliability at 100 \%. Audit pass rates stay within two cells of the full system. Collapsing all four agents into one fused context (B) does the same. Arm B built 12 of 12 with a clean audit on every model, the joint best result in the corpus, at 4.9 min against 6.1. The one arm-F model that failed the post-hoc audit is the cost of removing the gate. Without the gate, that model would have consumed solver time and entered the corpus as a result. The defensible claim for the decomposition is therefore not that it builds better models. It makes the build inspectable, localises failure to a role, and separates the checker from the builder.

\textbf{4. Surviving models agree exactly, not merely closely.} Of the 111 built cells with a comparable mesh fingerprint, 107 match the reference element count to the element. All four mismatches are the same problem across four arms (C, H3, P1, P2). Problem F1 was built at 97 812 elements against the correct 138 258. That is the signature of an omitted surcharge key, so the 20 kN/m² load was never applied. The settlement results show the same pattern. Of 74 solved cells comparable with the reference, 70 match to 0.00 \% in both maximum settlement and trough width. Three of the four exceptions are the extraction defects already named. The interpretation is stronger than ``settlement is insensitive to architecture''. Every correctly extracted specification produced an identical model, and every observed disagreement traces to a value read wrongly at intake. This should be read against the fact that all arms drive the same builder scripts from their extracted configuration. What varies between arms is the reasoning that writes the configuration, not the code that consumes it.

Accuracy against the manual expert is reported separately. The concordance reference is itself produced by the system under test, so it can only establish that the arms agree with one another. Figure 16 sets every solved arm against arm A over 86 comparable cells. Of these, 83 fall within 1.01 \% of the manual model in \(S_{max}\) and 1.21 \% in \(k\). Three do not. Arm C on a tier-B problem deviates by 12.0 \%, which is the extraction defect of finding 4. Arms C and P2 on problem F1 both deviate by −5.5 \%. The latter two are one finding rather than two. They were built on different days from independent extractions, and their maximum settlements agree to about one part in \(10^{16}\). Agreement at machine precision means the two extractions produced the same input deck. Problem F1 therefore admits two reproducible readings, giving \(S_{max}\) = 10.06 mm and 9.48 mm. The second is a property of the problem statement rather than of either arm. The full system reproduces the expert to a mean absolute deviation of 0.39 \% in \(S_{max}\) and 0.38 \% in \(k\). Four of twelve cells agree on \(k\) to three decimal places.

\begin{figure*}[!t]
\centering
\includegraphics[width=\textwidth,height=0.48\textheight,keepaspectratio]{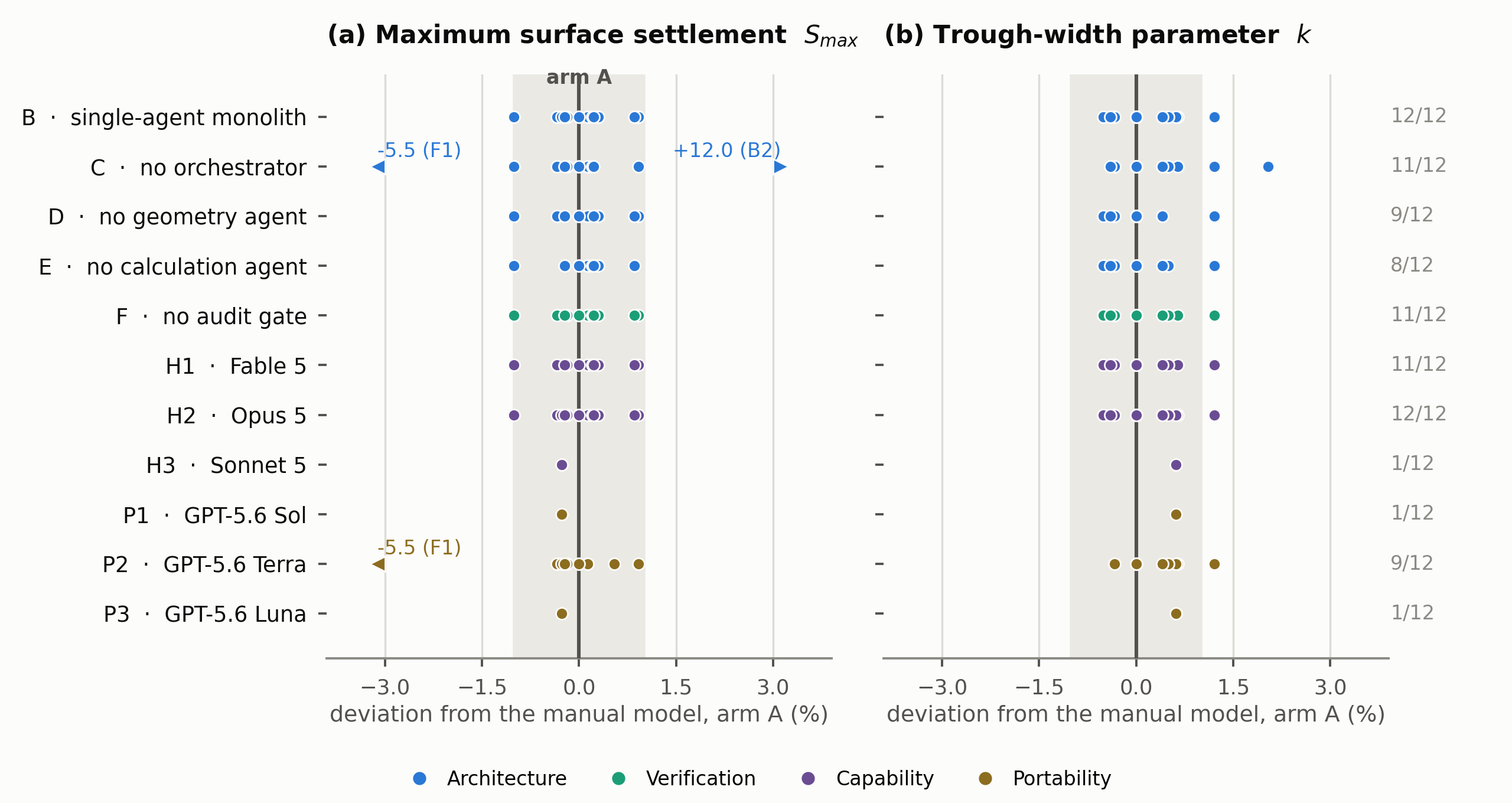}
\caption{Signed deviation of every solved arm from the manual expert model, arm A, over the twelve problems. Each mark is one problem. The zero line is the manual model and the shaded band is ±1 \%. Cells outside the axis are clipped at the margin with the value printed above the row.}
\label{fig:16}
\end{figure*}

Build time separates the arms in the opposite direction from robustness. Removing verification entirely (F) gives the fastest builds at 3.4 min. The full system pays a 2.7 min premium for the checkpoint and gate. That is under three minutes per model for an independent, information-barriered audit record. The scripted calculation arm (E) is slowest at 8.1 min, because its fixed staging recipe cannot adapt to the case. Cross-vendor portability is demonstrated but not at parity. The Codex arms reach 50 to 92 \% build reliability against the same server and skill files. Three of their nine halts are the drawdown convention splits that also account for one arm-G and three arm-H4 cells. This points to a harness or prompt-adherence gap in reading the skill modules, not to a limitation of the protocol.

\section{Field Application, Discussion and Limitations}\label{field-application-discussion-and-limitations}

\subsection{MRT Orange Line stacked twin-tunnel section}\label{mrt-orange-line-stacked-twin-tunnel-section}

The benchmark exercises the framework on idealised specifications. This section applies it to a real alignment: section G013B of the MRT Orange Line (East Section). This is a stacked twin-tunnel reach driven by EPB shields of 6.58 m shield diameter and 9.615 m length, with segmental rings of 5.7 m internal diameter and 0.3 m thickness (Duangsano et al., 2023). Both tunnels share the vertical plane at \(x = 0\). The first drive is at an axis depth of 34.0 m. The second, driven later, is at 17.0 m directly above it. The troughs therefore superpose on a common centreline, and the upper tunnel is driven through ground already disturbed by the lower one. The full four-agent configuration with all 15 modules and Opus 5 is to be used unchanged, with no repair loop and no manual correction. The section will then report what the deployed configuration produces, not what it can be tuned to produce. At the time of writing the field record has been assembled and the model has not yet been built.

The monitoring array was read after the first drive had passed and again after the second, so the record separates the two contributions (Fig. 17). Gaussian fits give \(S_{max}\) = 20.82 mm with \(i\) = 13.76 m and \(k\) = 0.40 for the first drive. The second drive adds \(S_{max}\) = 8.78 mm with \(i\) = 11.80 m and \(k\) = 0.69. The completed trough reaches about 30 mm at the centreline. The volume losses are 2.10 \% and 0.76 \%. Two features are what the automated model has to reproduce. The deep first drive gives the \(k\) = 0.40 that Bangkok practice reports for tunnels in the sand layer. The second drive, at half that depth, gives a markedly wider \(k\) = 0.69 relative to its own axis depth. The first drive's volume loss is near the upper end of the range recorded on the project, and the second is well inside it.

\begin{figure}[!t]
\centering
\includegraphics[width=\columnwidth]{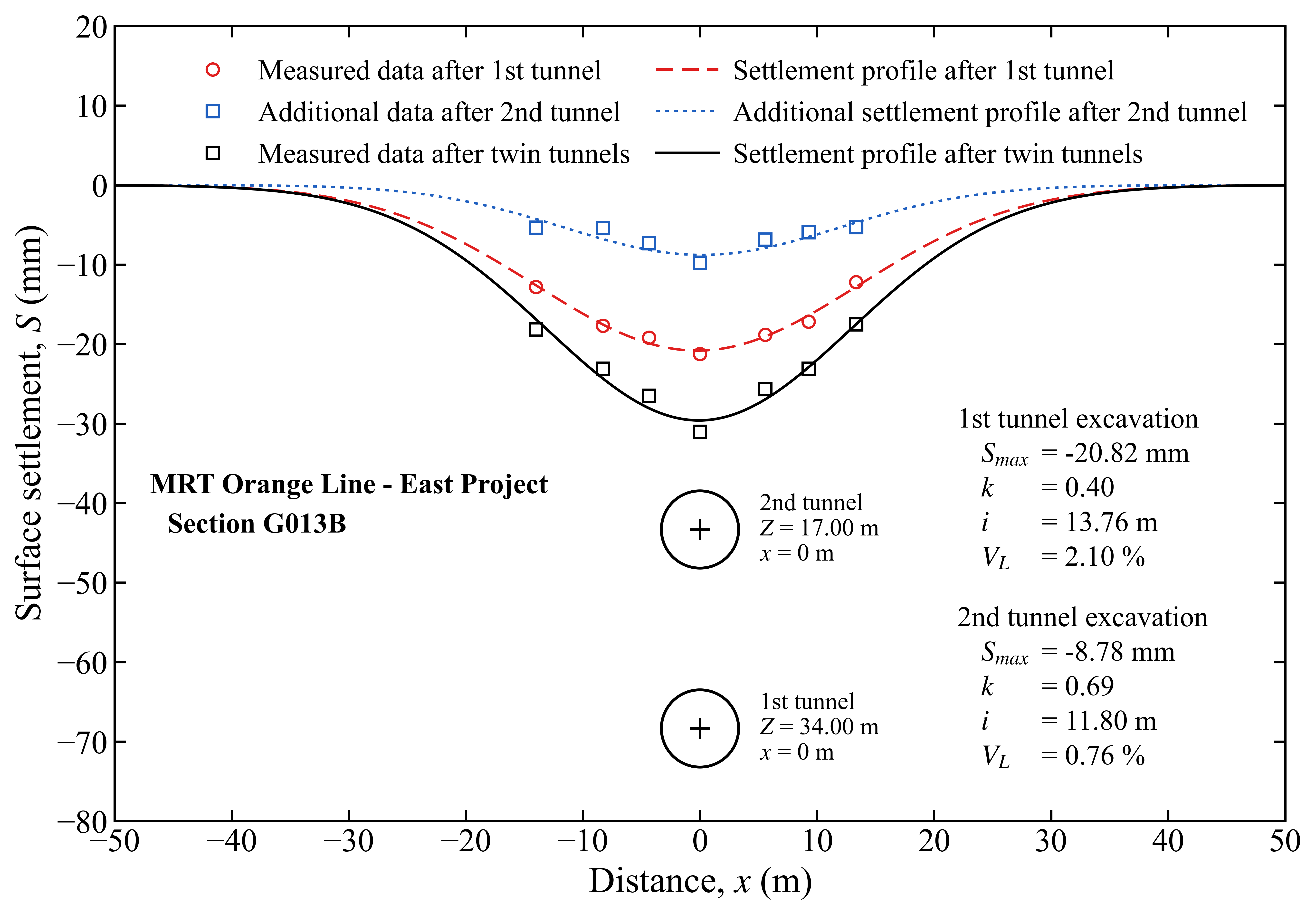}
\caption{Measured surface settlement at section G013B: the trough after the first (lower) drive, the additional settlement after the second (upper) drive, and the completed twin-tunnel trough, each with its fitted Gaussian.}
\label{fig:17}
\end{figure}

\begin{quote}
{[}TO BE ADDED: section chainage, stratigraphy at axis level and groundwater condition; recorded face pressure, grout pressure, contraction and advance rate per drive as applied in the model, the lagging distance, and the element count and build time; the automated predictions overlaid on these arrays, with a table of predicted against measured \(S_{max}\), \(i\), \(k\) and \(V_L\) and the percentage deviation.{]}
\end{quote}

\subsection{Comparison against published Bangkok ranges}\label{comparison-against-published-bangkok-ranges}

Duangsano et al.~(2023) report 18 settlement arrays from the Orange Line combined with 49 from the Blue Line. This gives 67 monitored cross-sections with a tunnel geometry closely matching benchmark tunnel 1. Most troughs are well fitted by a Gaussian. Most volume losses lie between 0.5 \% and 1.0 \%, with an average of 0.75 \%. Individual drives span 0.15 to 2.23 \%. The trough width follows \(i = kZ\) with \(k \approx 0.5\) for clayey soils and \(k \approx 0.4\) in sand. Suwansawat and Einstein (2007) place most second-tunnel data between \(i = 0.4Z\) and \(0.6Z\). Table VI sets the automated predictions against these ranges.

\begin{table*}[!t]
\caption{Automated benchmark predictions against measured Bangkok ranges.}
\label{tab:6}
\centering\footnotesize
\setlength{\tabcolsep}{4pt}\renewcommand{\arraystretch}{1.15}
\begin{tabularx}{\textwidth}{YYYY}
\toprule
Quantity & Automated benchmark (Table IV) & Bangkok field range & Verdict \\
\midrule
$S_{max}$, single tunnel & 3.5–14.9 mm & 5–30 mm twin-tunnel totals, single-drive components 3.2–21.3 mm & consistent \\
$V_L$ & 0.51–1.24 \%, mean 0.87 \% (11 of 12 within 0.5–1.0 \%) & most drives 0.5–1.0 \%, mean 0.75 \% (spread 0.15–2.23 \%) & consistent \\
$k$, axis in clay & 0.44–0.49, mean 0.47 & ≈ 0.5 (range 0.4–0.6) & consistent \\
$k$, axis at clay/sand interface & 0.49–0.50, mean 0.50 & 0.4–0.6 & consistent \\
$k$, axis in sand & 0.59–0.61, mean 0.59 & ≈ 0.4 (range 0.3–0.5) & over-predicts trough width \\
\bottomrule
\end{tabularx}
\end{table*}

Three points follow. \textbf{First}, in clay and at the clay/sand interface, which covers eight of the twelve problems, the automated models land inside the measured envelope on all three quantities. No calibration to field data entered the pipeline. The stiff-clay case H2\_A2 sits almost exactly on the measured central values. \textbf{Second}, volume loss is consistent throughout, but only once the reference diameter is made explicit. The models reference \(V_L\) to the shield diameter. Duangsano et al.~(2023) define it with ``D = tunnel diameter'' without saying which. Back-calculating their Table 3 reproduces every printed \(V_L\) to ±0.01 \% using the 6.58 m shield diameter, so the two are on the same basis. The distinction is not negligible. Substituting the lining diameter inflates the computed volume loss by 4 to 11 \%, depending on the tunnel. \textbf{Third}, and negatively, the four sand-axis cases systematically over-predict trough width. They give \(k \approx 0.59\) where practice indicates \(k \approx 0.4\). These cells reproduce the reference mesh element for element, so the specification was extracted correctly. The discrepancy is constitutive rather than extractive. The most plausible cause is the Hardening Soil stiffness assigned to the sand and the absence of small-strain stiffness, which is known to widen and shallow computed troughs. A rerun of the four sand-axis cases with a small-strain stiffness model is the direct test, and it has not yet been made. This is exactly the systematic bias that an audit against the specification cannot detect. The model faithfully implements what it was asked to build. The result delimits what the verification architecture provides. The gate certifies that the model matches the specification. It does not certify that the specification produces the right physics.

\subsection{Discussion}\label{discussion}

\textbf{Why domain skills dominate.} Two conventions account for almost every halt. The first is material-set construction and assignment, which caused 11 of 12 arm-G halts. The obstacle is not the command. It is knowing which nine strata exist, which parameter set each takes, and the exact object names the project's material sets carry. The second is the drawdown convention splits. These caused the remaining arm-G cell, three arm-H4 cells, and three of the nine Codex halts, where they were the most common named cause. Without splitting the profile at the midpoint band, the regime cannot be represented at all. Both are project conventions with mechanical consequences. They are not derivable from first principles, and they are not in general pre-training corpora. Getting them wrong produces no model rather than a slightly worse one. That is what the skill modules encode, and it explains the shape of the result. The effect is total rather than graded because the knowledge is procedural and binary in its consequences. Two observations sharpen this. A knowledge-ablated agent does not always fail fast. Cell G\_A1 ran 127.6 minutes of plausible-looking work. When failure is slow, only an independent checker stops it propagating into a solved model. And the same convention that defeats a knowledge-ablated frontier model also defeats capable models on other harnesses with the skill files present. The operative variable is therefore whether the procedural knowledge reaches the agent's working context, not whether it exists on disk.

\textbf{Why the decomposition did not pay off where expected.} The likely mechanism for finding 3 is a handoff-loss effect. Every disagreement in the corpus originated at intake rather than in construction. A decomposed pipeline offers little protection against that class of defect. What crosses each role boundary is the configuration, not the reasoning that produced it. A key never written is indistinguishable downstream from a key deliberately left at its default. The omitted surcharge key on F1 is the clearest instance. It reached four arms spanning both harnesses and two architectures. In a fused context, the same reasoning that reads the specification also writes the configuration. A doubt raised while reading cannot be dropped at a handoff. That fusion also removes the independent reader, so it is a trade rather than a gain. Two qualifications keep this from being an argument against decomposition as such. The benchmark problems sit within a single frontier model's context and competence. The handoff cost is real, but the context-partitioning benefit has nothing to bite on. Twin-tunnel and multi-drive problems are where that benefit would be expected to appear. And reliability is not all the decomposition buys. It buys attributable failure, which is what made the taxonomy above possible, and a checker that does not share the builder's assumptions. On this benchmark its value is diagnostic and evidentiary rather than generative. A study reporting only end-to-end success rates could not have said so either way.

\textbf{Computational cost.} The dominant cost is not the agent but the solver. Build time is 3.4 to 8.1 minutes per model, against solver times measured in hours. A saved pre-calculation model of 200 to 300 MB expands to several gigabytes once solved. This reverses the conventional bottleneck. Expert-hours removed from construction are replaced by queue time on the solve machine. On tokens, composition matters. Across round 1 the Claude arms billed 61.5 M tokens (11.3 M output, 50.1 M cache creation), and a further 556.5 M were read from cache. Cache reads are therefore 91.7 \% of everything the agents read. An estimate that treats all input alike overstates the bill by an order of magnitude. An estimate built on fresh input alone (0.2 M over the whole corpus) understates it by two. The two cost measures also do not rank the arms alike. The monolith is the fastest surviving Claude arm but the most expensive on tokens. Against expert time, paired on the same twelve problems, the totals are 296 min by hand (mean 24.6, SD 4.8) and 73 min automated (mean 6.1, SD 3.7). The ratio of means is 4.1 and the paired median is 4.8. The difference is significant on a Wilcoxon signed-rank test (p = 0.0005), and the automated build was faster on all twelve problems. Two qualifications apply. The comparison is build time only. It omits the specification reading before a manual build, the review that should follow an automated one, and solver time. The narrowest ratio, 1.7 on A1, is the system's cold start (16.4 min against 3.3 min on the next problem), so the steady-state ratio is nearer the median.

\textbf{Practical implications.} The framework is an accelerant, not a replacement. The responsible engineer remains the author of the specification and the consumer of the audited result. The tagged intake and the independent audit produce an inspectable trail from specification to settlement that a conventional GUI session does not.

\subsection{Limitations and threats to validity}\label{limitations-and-threats-to-validity}

\begin{itemize}
\tightlist
\item
  \textbf{Replication is incomplete.} Round 1 gives n = 1 per cell. The intervals in Table V quantify uncertainty across problems, not run-to-run variance, and an LLM pipeline is stochastic. Rounds 2 and 3 will bring every automated cell to n = 3. This is the single largest limitation. It bears most heavily on arms that differ by one or two cells. It does not threaten the two large effects, arms G and H4 at 0 of 12 against every other Claude arm at 12 of 12.
\item
  \textbf{The concordance reference is not independent, and is not asked to be.} The mesh fingerprint establishes consistency between arms, not accuracy. That gap is closed from outside by arm A. The external check does not cover run-to-run variance, because arm A is a single manual build per problem.
\item
  \textbf{Two claimed contributions are not isolated by an arm.} The information barrier and the tagged intake had dedicated arms under a prior protocol. Both were dropped when the roster was frozen. Arm B provides indirect evidence on the barrier, but neither contribution is measured here.
\item
  \textbf{A harness asymmetry exists.} The agent CLI changed its exposed tool names mid-campaign. Arms H2, C, D, E, F, G and H1 ran with one shell tool listed, and arms H3, H4 and B with another. One spurious total failure was caused, detected and re-run. The arms were not executed under a byte-identical harness. Arm H4's total halt deserves confirmation under the corrected configuration.
\item
  \textbf{Context contamination is the dominant internal-validity threat.} It is a harness-level threat, not a prompt-level one. Mid-campaign we found that the harness injected the project's orientation file and memory index into every subagent before it read its task prompt. No agent-level instruction could suppress this. It is fatal to the knowledge-ablation arm specifically. With the injection path open, ``skills removed'' is not enforceable. A weaker rung can source a specification value from injected memory while sincerely tagging it as read from the source. This behaviour was directly observed. The entire affected corpus was discarded. The extraction workspace was relocated outside the project root, and every subsequent harness change was gated with an automated injection probe. Scans over round 1 report zero breaches. Readers should treat the probe, not the instructions given to the agents, as the operative control. This is the principal methodological lesson of the study.
\item
  \textbf{Authorship independence is partial.} The skill modules and audit configurations are authored by the same team. Each case's audit configuration is transcribed from the source documents by an extraction step that is itself an agent. The safeguard is that the auditor recomputes every convention independently and cannot import conversion logic from the build scripts. Its own sensitivity is uncharacterised. Until the planned fault-injection study runs, its false-negative rate is unknown, and that bounds what the verification contribution can claim. Independence of the accuracy reference holds by construction. Arm A was built with no agent, no MCP server and no skill module in the loop.
\item
  \textbf{Scope.} Two idealised soil profiles are used, and real stratigraphy is laterally variable. The domain depth of 50 m leaves about 1.5 shield diameters below the invert of the large tunnel in tier C, and the influence of the bottom boundary on those troughs has not been checked. Twin-tunnel and lagging-distance problems are outside the scored matrix. The paired field comparison is confined to one Orange Line section, which is a demonstration rather than a validation set. Third-vendor portability (P4) is unexecuted. Results are tied to specific LLM versions and may drift. The rungs are ordered by vendor tier rather than by a measured agentic score. A single FE package and problem domain are covered. ``Build reliability'' as defined is a single unaided attempt with no repair budget. This is a deliberately conservative construct, and it understates what the deployed system achieves with an engineer in the recovery role.
\end{itemize}

\section{Conclusions}\label{conclusions}

This paper presented a multi-agent framework in which a large language model drives PLAXIS 3D end-to-end, from a natural-language specification to an extracted settlement trough. The route is a purpose-built MCP server layered on the official remote-scripting interface. The main conclusions are:

\begin{enumerate}
\def\labelenumi{\arabic{enumi}.}
\tightlist
\item
  \textbf{Feasibility.} With no human operating the graphical interface, the system performs project set-up, stratigraphy, Bangkok hardening-soil materials, TBM and lining construction, meshing, groundwater assignment, ring-by-ring staged advance, independent audit, calculation and settlement extraction through 19 MCP tools. It built 12 of 12 models at a mean of 6.1 minutes with a clean audit on every one. The same models took 24.6 minutes by hand.
\item
  \textbf{Domain knowledge is the decisive component.} With the complete architecture, the full tool set and the strongest base model but the 15 modules withheld, no model was produced. Failures concentrate on two project conventions: Bangkok material-set construction and the drawdown-band layer splits. Both are procedural, absent from general pre-training, and binary in their consequences. Architecture cannot substitute for them.
\item
  \textbf{Base-model capability behaves as a threshold, not a gradient.} Fable 5, Opus 5 and Sonnet 5 each built 12 of 12 under an identical architecture and skill set. Haiku 4.5 halted on all 12 within 0.3 minutes on average, on specification reading rather than modelling. The rungs are ordered by vendor tier, so the threshold can only be bracketed between the Haiku and Sonnet tiers. It cannot be located on an external scale. These findings rest on one run per cell and await the replication rounds.
\item
  \textbf{The decomposition's value is diagnostic, not generative.} This is a negative result reported as found. Removing the orchestrator, either builder agent, or the entire verification role left build reliability unchanged. Collapsing all four into one context was as reliable and faster. What the decomposition demonstrably provides is attributable failure and a checker that does not share the builder's assumptions. Only a component-level ablation could have established this.
\item
  \textbf{Accuracy.} Among models that build, agreement is exact rather than approximate. 107 of 111 comparable cells reproduce the concordance mesh to the element, and 70 of 74 solved cells reproduce its settlement to 0.00 \%. Against the external reference, 83 of 86 solved cells agree within 1.01 \% in \(S_{max}\) and 1.21 \% in \(k\). The full system agrees to a mean absolute 0.39 \% and 0.38 \%. The representative case yields \(S_{max}\) = 11.6 mm, \(i\) = 9.8 m, \(k\) = 0.49 and \(V_L\) = 0.84 \%, within reported Bangkok MRT ranges. The tier-A sequence reproduces the expected settlement-depth and trough-width trends without having been given them.
\item
  \textbf{Extraction, not construction, is where the remaining error lives.} Every disagreement originated at intake. Where a key is absent, an unguarded default in the shared builder script can convert the omission into a model that looks well-formed. This relocates the reliability problem from code generation to specification reading and defensive tool design.
\item
  \textbf{Methodological lesson, and portability.} An earlier corpus was discarded in full because the harness auto-injected project context into every subagent. That made the knowledge-ablation arm unenforceable. Any ablation that claims to remove a knowledge source must demonstrate the removal at the harness level with an automated probe, not by instructing the agent. The same server design also supports other vendors. The three GPT-Codex arms reached 50 to 92 \% build reliability against identical skill files. Third-vendor portability, replication rounds 2 and 3, twin-tunnel problems and the paired Orange Line comparison are ongoing.
\end{enumerate}

For practice, the implications are threefold. A parametric study whose cost was dominated by expert modelling hours can now be run at roughly a quarter of the build time per model. The expert is out of the loop between specification and audited result, so the bottleneck moves to solver throughput. The independent audit produces an inspectable, per-check trail from specification to settlement that a conventional GUI session does not. That trail is the artefact a reviewing engineer, checker or regulator actually needs. And because the domain knowledge lives in fifteen versionable files rather than in a prompt or in fine-tuned weights, a firm's own conventions can be encoded, inspected and maintained the way code is. Their contribution to reliability can then be measured the way this study measured Bangkok's.

\section*{References}\label{references}

\begingroup\footnotesize\setlength{\parindent}{-1.2em}\setlength{\leftskip}{1.2em}\setlength{\parskip}{2pt}

\par
Anthropic (2024). Model Context Protocol: an open standard for connecting AI applications to external systems. https://modelcontextprotocol.io
\par
Bekele, Y. W. (2025). GeoSim.AI: AI assistants for numerical simulations in geomechanics. arXiv:2501.14186.
\par
Bekele, Y. W. (2026). GeoMCP: a trustworthy framework for AI-assisted analytical geotechnical engineering. arXiv:2603.01022.
\par
Bentley Systems. \emph{PLAXIS 3D Reference Manual: remote scripting server and plxscripting API}. Bentley Systems, Exton, PA.
\par
Chatterjee, K., Desai, M., \& Li, J. (2026). Application of large language models in geotechnical engineering: a movement towards safe and sustainable future. \emph{Geotechnics}, 6(2), 38. https://doi.org/10.3390/geotechnics6020038
\par
Duangsano, O., Kandavorawong, K., Asanprakit, A., \& Jongpradist, P. (2023). Ground movements in shield tunnelling for MRT Orange Line in Bangkok. In \emph{Geotechnics for Sustainable Infrastructure Development --- Geotec Hanoi 2023} (Phung, Ed.), paper 558. https://doi.org/10.1007/978-981-99-9722-0\_37
\par
Ekaksorn, N., et al.~(2024). Comparative evaluations of building impact assessment due to adjacent tunneling. \emph{IOP Conf. Series: Earth and Environmental Science}, 1333, 012052. https://doi.org/10.1088/1755-1315/1333/1/012052
\par
Elbasani, E., Tran, T. B., Lee, S., \& Lee, J. (2026). LLM-enabled automation of CAD--structural analysis via Model Context Protocol. \emph{Automation in Construction}, 188, 107030. https://doi.org/10.1016/j.autcon.2026.107030
\par
Fan, E., Hu, K., Wu, Z., Ge, J., Miao, J., Zhang, Y., Sun, H., Wang, W., \& Zhang, T. (2025). ChatCFD: an LLM-driven agent for end-to-end CFD automation with structured knowledge and reasoning. arXiv:2506.02019.
\par
Fan, L., Liu, F., \& Chen, C. (2026). Domain adaptation of large language models for geotechnical applications. \emph{Solid Earth Sciences}, 11, 100285. https://doi.org/10.1016/j.sesci.2025.100285
\par
Geng, Z., Franklin, I., Martinez, S., Liu, J., Zhao, Y., \& Cheng, M. (2026a). Agentic large language models for automated structural analysis of 3D frame systems. arXiv:2606.06525.
\par
Geng, Z., Liu, J., Franklin, I., Cao, R., Frangopol, D. M., \& Cheng, M. (2026b). Automating structural analysis across multiple software platforms using large language models. arXiv:2604.09866.
\par
Geng, Z., Liu, J., Cao, R., Cheng, L., Frangopol, D. M., \& Cheng, M. (2026c). A novel multi-agent architecture to reduce hallucinations of large language models in multi-step structural modeling. arXiv:2603.07728.
\par
Heama, N., Jongpradist, P., Lueprasert, P., Suwansawat, S., \& Jamsawang, P. (2021). Comparative effects of adjacent loaded pile row on existing tunnel by 2D and 3D simulation models. \emph{Geomechanics and Engineering}, 27(2), 151--165. https://doi.org/10.12989/gae.2021.27.2.151
\par
Heimig-Elschner, T., Du, C., Scheuvens, A., Borrmann, A., \& Beetz, J. (2026). A modular reference architecture for MCP-servers enabling agentic BIM interaction. arXiv:2601.00809.
\par
Hu, M., Gao, H., Mi, Q., Wu, B., Lu, J., \& Liu, Y. (2025). Bridging the information gap in smart construction: an LLM-based assistant for autonomous TBM tunneling. \emph{Smart Cities}, 8(6), 212. https://doi.org/10.3390/smartcities8060212
\par
Liang, H., Zhou, Y., Talebi-Kalaleh, M., \& Mei, Q. (2025). Automating structural engineering workflows with large language model agents. arXiv:2510.11004.
\par
Likitlersuang, S., Surarak, C., Suwansawat, S., Wanatowski, D., Oh, E., \& Balasubramaniam, A. S. (2014). Simplified finite-element modelling for tunnelling-induced settlements. \emph{Geotechnical Research}, 1(4), 133--152. https://doi.org/10.1680/gr.14.00016
\par
Loganathan, N., \& Poulos, H. G. (1998). Analytical prediction for tunneling-induced ground movements in clays. \emph{J. Geotechnical and Geoenvironmental Engineering}, 124(9), 846--856. https://doi.org/10.1061/(ASCE)1090-0241(1998)124:9(846)
\par
Lueprasert, P., Jongpradist, P., \& Suwansawat, S. (2017). Tunneling simulation in soft ground using shell elements and grouting layer. \emph{International Journal of GEOMATE}, 12(31), 51--57. https://doi.org/10.21660/2017.31.6535
\par
Lueprasert, P., Jongpradist, P., Jongpradist, P., \& Schweiger, H. F. (2023). Structural responses of a tunnel lining due to an adjacent loaded pile. \emph{International Journal of Civil Engineering}, 21, 1027--1043. https://doi.org/10.1007/s40999-023-00821-9
\par
Mair, R. J., \& Taylor, R. N. (1997). Bored tunnelling in the urban environment. \emph{Proc. 14th ICSMFE}, Hamburg, 2353--2385.
\par
O'Reilly, M. P., \& New, B. M. (1982). Settlements above tunnels in the United Kingdom --- their magnitude and prediction. \emph{Tunnelling '82}, London, 173--181.
\par
Pan, H., Chard, R., Mello, R., Grams, C., He, T., Brace, A., et al.~(2025). Experiences with Model Context Protocol servers for science and high performance computing. arXiv:2508.18489.
\par
Parsa-Pajouh, A. (2025). Application of generative AI to automate numerical analysis and synthetic data generation in geotechnical engineering. \emph{Machine Learning and Data Science in Geotechnics}, 1(1), 50--59. https://doi.org/10.1108/MLAG-09-2024-0008
\par
Peck, R. B. (1969). Deep excavations and tunneling in soft ground. \emph{Proc. 7th ICSMFE}, Mexico City, State-of-the-Art Volume, 225--290.
\par
Phutthananon, C., Lertkultanon, S., Jongpradist, P., Duangsano, O., Likitlersuang, S., \& Jamsawang, P. (2023). Numerical investigation on the responses of existing single piles due to adjacent twin tunneling considering the lagging distance. \emph{Underground Space}, 11, 171--188. https://doi.org/10.1016/j.undsp.2022.12.005
\par
Qi, Y., Xu, R., \& Chu, X. (2025). FeaGPT: an end-to-end agentic-AI for finite element analysis. arXiv:2510.21993.
\par
Sagaseta, C. (1987). Analysis of undrained soil deformation due to ground loss. \emph{Géotechnique}, 37(3), 301--320. https://doi.org/10.1680/geot.1987.37.3.301
\par
Sheil, B., Anagnostopoulos, C., Buckley, R., Ciantia, M. O., Febrianto, E., Fu, J., et al.~(2026). Artificial intelligence transformations in geotechnics: progress, challenges and future enablers. \emph{Computers and Geotechnics}, 189, 107604. https://doi.org/10.1016/j.compgeo.2025.107604
\par
Sittiamornporn, N., Phutthananon, C., Jongpradist, P., Chayaroon, A., \& Malaisree, P. (2024). Comparative effects of twin-stacked tunnel excavation on adjacent structure using different simulation models. \emph{ITA-AITES World Tunnel Congress 2024}, Shenzhen.
\par
Surarak, C., Likitlersuang, S., Wanatowski, D., Balasubramaniam, A., Oh, E., \& Guan, H. (2012). Stiffness and strength parameters for hardening soil model of soft and stiff Bangkok clays. \emph{Soils and Foundations}, 52(4), 682--697. https://doi.org/10.1016/j.sandf.2012.07.009
\par
Suwansawat, S., \& Einstein, H. H. (2006). Artificial neural networks for predicting the maximum surface settlement caused by EPB shield tunneling. \emph{Tunnelling and Underground Space Technology}, 21(2), 133--150. https://doi.org/10.1016/j.tust.2005.06.007
\par
Suwansawat, S., \& Einstein, H. H. (2007). Describing settlement troughs over twin tunnels using a superposition technique. \emph{J. Geotechnical and Geoenvironmental Engineering}, 133(4), 445--468. https://doi.org/10.1061/(ASCE)1090-0241(2007)133:4(445)
\par
Verruijt, A., \& Booker, J. R. (1996). Surface settlements due to deformation of a tunnel in an elastic half plane. \emph{Géotechnique}, 46(4), 753--756. https://doi.org/10.1680/geot.1996.46.4.753
\par
Xu, C., Cao, B. T., Yuan, Y., \& Meschke, G. (2024). A multi-fidelity deep operator network (DeepONet) for fusing simulation and monitoring data: application to real-time settlement prediction during tunnel construction. \emph{Engineering Applications of Artificial Intelligence}, 133, 108156. https://doi.org/10.1016/j.engappai.2024.108156 (preprint arXiv:2310.00057)
\par
Xu, H.-R., Zhang, N., Yin, Z.-Y., \& Atangana Njock, P. G. (2025). GeoLLM: a specialized large language model framework for intelligent geotechnical design. \emph{Computers and Geotechnics}, 177, 106849. https://doi.org/10.1016/j.compgeo.2024.106849
\par
Ye, C., Xiong, G., Shang, J., Dai, X., Gong, X., \& Lv, Y. (2025). SUMO-MCP: leveraging the Model Context Protocol for autonomous traffic simulation and optimization. arXiv:2506.03548.
\par
Youwai, S., Phim, D., Murcia, V. G., \& Onas, R. C. (2026). Large language model-based multi-agent systems for automated foundation design: router-driven task classification and expert selection framework. \emph{AI in Civil Engineering}, 5, 5. https://doi.org/10.1007/s43503-026-00088-8
\par
Yue, L., Somasekharan, N., Zhang, T., Cao, Y., \& Pan, S. (2025). Foam-Agent 2.0: an end-to-end composable multi-agent framework for automating CFD simulation in OpenFOAM. arXiv:2509.18178.
\par
Zhang, J., Lao, J., Liu, C., Liu, S., Wu, S., Zhang, L., Wang, B., \& Huang, S. (2026). VFEAgent: a multimodal agent framework for end-to-end automated finite element analysis. arXiv:2605.28978.
\par\endgroup

\end{document}